\documentclass[reprint,twocolumn]{revtex4}
\usepackage{amsfonts}
\usepackage{amsmath}
\usepackage{amssymb}
\usepackage{charter}
\usepackage{graphicx}
\usepackage{float}
\usepackage{amsmath}
\usepackage{amssymb}
\usepackage{charter}
\usepackage{graphicx}
\usepackage{subfigure}
\usepackage{graphicx}
\usepackage{epstopdf}
\usepackage{color}
\usepackage{tocvsec2}
\usepackage{enumerate}
\usepackage{graphicx}
\usepackage{dcolumn}
\usepackage{bm}

\begin{document}

\title{Improved linear and Kerr nonlinear phase estimation via photon
addition operations}
\author{Zekun Zhao$^{1}$}
\author{Qingqian Kang$^{1,2}$}
\author{Shoukang Chang$^{3}$}
\author{Teng Zhao$^{1}$}
\author{Cunjin Liu$^{1}$}
\author{Xin Su$^{1}$}
\author{Liyun Hu$^{1,4}$}
\thanks{hlyun@jxnu.edu.cn}
\affiliation{$^{{\small 1}}$\textit{Center for Quantum Science and Technology, Jiangxi
Normal University, Nanchang 330022, China}\\
$^{{\small 2}}$\textit{Department of Physics, Jiangxi Normal University
Science and Technology College, Nanchang 330022, China}\\
$^{{\small 3}}$\textit{MOE Key Laboratory for Nonequilibrium Synthesis and
Modulation of Condensed Matter, Shaanxi Province Key Laboratory of Quantum
Information and Quantum Optoelectronic Devices, School of Physics, Xi'an
Jiaotong University, Xi'an 710049, People's Republic of China}\\
$^{{\small 4}}$\textit{Institute for Military-Civilian Integration of
Jiangxi Province, Nanchang 330200, China}}

\begin{abstract}
The accuracy of quantum measurements can be effectively improved by using
both photon-added non-Gaussian operations and Kerr nonlinear phase shifters.
Here, we employ coherent state mixed photon-added squeezed vacuum state as
input into a Mach-Zehnder interferometer with parity detection, thereby
achieving a significant enhancement in phase measurement accuracy. Our
research focuses on phase sensitivity of linear phase shift under both ideal
conditions and photon loss, as well as quantum Fisher information. The
results demonstrate that employing the photon addition operations can
markedly enhance phase sensitivity and quantum Fisher information, and the
measurement accuracy can even approach the Heisenberg limit. In addition, we
delve deeper into the scenario of replacing the linear phase shifter with a
Kerr nonlinear one and systematically analyze the quantum Fisher information
under both ideal and photon loss conditions. By comparison, it is evident
that employing both the photon addition operations and the Kerr nonlinear
phase shifter can further significantly enhance phase measurement accuracy
while effectively improving the system's robustness against photon loss.
These findings are instrumental in facilitating the development and
practical application of quantum metrology.

\textbf{PACS: }03.67.-a, 05.30.-d, 42.50,Dv, 03.65.Wj
\end{abstract}

\maketitle

\section{Introduction}

Quantum metrology leverages the non-classical characteristics of quantum
states to achieve high-precision measurements, with broad applications in
various fields of science and technology \cite{1,2,3,4,5,6}. According to
the principles of quantum mechanics, it is theoretically possible to surpass
the accuracy limit set by classical physics, known as the standard quantum
limit (SQL). This limit is represented by $1/\sqrt{\bar{N}}$, where $\bar{N}$
is the total average photon number in the quantum state input to the system
to be measured \cite{7,8,9}. In the early 1980s, Caves introduced a coherent
state (CS) and a squeezed vacuum state (SVS) into the two input ports of the
Mach-Zehnder interferometer (MZI), enabling the measurement accuracy to
surpass the SQL due to the non-classical properties of the input quantum
states \cite{9}.

The non-classical characteristics of quantum states, particularly
entanglement, have been identified as valuable resources for quantum
technologies such as quantum communication \cite{10,11}, quantum key
distribution \cite{12,13,14}, and quantum metrology \cite%
{15,16,17,18,19,20,21}. In quantum metrology, widely used interferometers
comprise the linear MZI and the nonlinear SU(1,1) interferometer \cite{22}.
Numerous quantum states, such as NOON states \cite{15,16,17,18} and
entangled CSs \cite{19,20,21}, can be used as inputs to these
interferometers, enhancing measurement accuracy beyond the SQL and
potentially surpassing the Heisenberg limit (HL) of $1/\bar{N}$ \cite{23}.
However, these states are not only more challenging to prepare but also
exhibit heightened sensitivity to noise. For instance, while theoretically,
highly entangled NOON states as inputs can achieve measurement accuracies up
to the HL, experimental preparation is difficult and unstable \cite%
{18,24,25,26,27}. Thus, the preparation of non-classical quantum states,
along with their robust anti-noise capabilities and stable transmission
properties, holds significant importance in the field of quantum metrology.
Recently, significant attention has been paid to the preparation of quantum
states via non-Gaussian operations and their applications in quantum
precision measurements. It has been demonstrated that non-Gaussian
operations such as photon subtraction, photon addition, and photon catalysis
effectively enhance non-classicality \cite{28,29}, thereby improving
measurement accuracy \cite{30,31,32,33,34,35,36,37,38}. For example, Wang
\emph{et al.} proposed utilizing CS combined with photon-added SVS (PASVS)
as inputs for lossless MZI to enhance measurement accuracy and compared it
with CS mixed with photon-subtracted SVS (PSSVS) inputs. The results suggest
that, under ideal conditions, photon addition and subtraction significantly
enhance the phase sensitivity of parity detection. Notably, photon addition
offers superior advantages in increasing phase measurement precision \cite%
{38}.

To improve measurement accuracy more effectively, replacing a linear phase
shifter with a nonlinear one in conventional interferometers has gained
increasing interest. Boixo \emph{et al.} demonstrated that when employing
entangled and product states as inputs to $k$th-order nonlinear
Hamiltonians, the limits of phase sensitivity can reach $\bar{N}^{-k}$ and $%
\bar{N}^{-\left( k-1/2\right) }$ respectively \cite{39,40}. In addition,
recent studies have suggested schemes to improve the precision using a Kerr
nonlinear phase shifter, showing that such approaches can surpass the HL
\cite{21,41,42,43,44,45,46,47}. For example, Zhang \emph{et al.} proposed
enhancing the precision in an MZI by replacing the linear phase shifter with
a Kerr nonlinear one, and used CS as the input state and parity detection.
The results indicated that the parity detection signal offers high
super-resolution for Kerr nonlinear phase shift, and the phase sensitivity
effectively surpasses the HL, and approaches quantum Cram\'{e}r-Rao bound
(QCRB) \cite{47}. However, the approach of employing both non-Gaussian
operations and Kerr nonlinear phase shifter to improve phase estimation
remains in the exploratory stage. Furthermore, photon loss unavoidably
impacts measurement accuracy in practical applications, and there is a
significant lack of systematic research addressing photon loss in this
context.

Inspired by previous studies, this paper proposes the scheme of using CS and
PASVS inputs to MZI and employing parity detection to achieve an effective
improvement in the measurement accuracy. By examining the phase sensitivity
and the quantum Fisher information (QFI) in both ideal and photon loss
scenarios, it is evident that the non-Gaussian operations of photon addition
significantly enhance the phase sensitivity and QFI. By increasing in the
addition photon number, the phase sensitivity and QCRB can be significantly
improved and can effectively outperform SQL and even approach HL.\
Furthermore, by considering the effects of photon loss, it can be found that
the photon addition operations can effectively enhance the robustness of the
system. We further replace the linear phase shifter with a Kerr nonlinear
one in the MZI and investigate the QFI. By comparing the linear phase shift
(under ideal conditions and photon loss), it is evident that the utilization
of Kerr nonlinear phase shifter combined with photon addition can
substantially enhance the QCRB. Moreover, the QCRB not only surpasses the
sub-Heisenberg limit (sub-HL) of $1/\bar{N}^{3/2}$, but also potentially
reaches the super-Heisenberg limit (SHL) of $1/\bar{N}^{2}$.

This paper is structured as follows: In Section II, we propose a phase
estimation model that utilizes a combination of CS and PASVS as the input
state for the MZI, incorporates parity detection at the output, and accounts
for photon loss. Section III delves into the phase sensitivity in both the
ideal case and the photon loss case. Section IV focuses on QFI including the
effects of photon loss. Section V further explores the scheme of improving
QFI and QCRB using Kerr nonlinear phase shift. Lastly, a summary is
presented in the concluding section.

\section{Phase estimation model for CS mixed PASVS input MZI}

\begin{figure}[tph]
\label{Fig1} \centering \includegraphics[width=0.83\columnwidth]{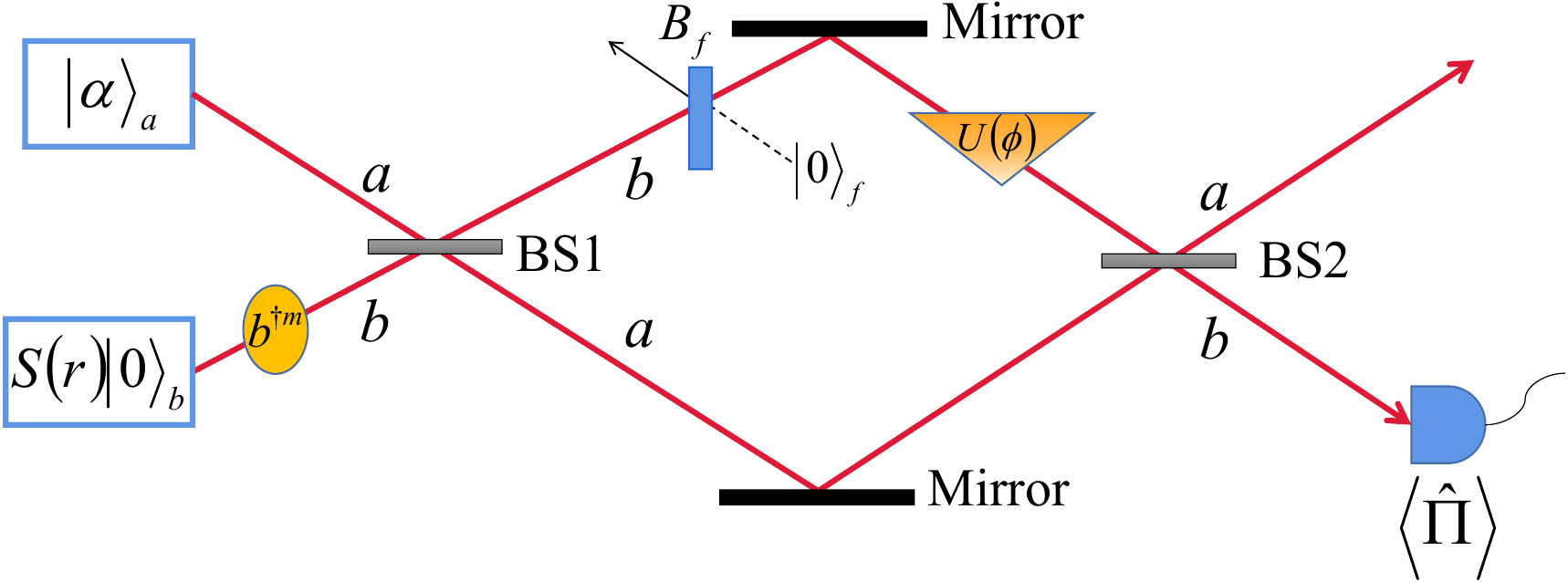}
\caption{Schematic diagram of phase estimation using CS and PASVS inputs in
an MZI. In this configuration, photon loss within the MZI is simulated by a
virtual beam splitter, and a parity detection scheme is employed at the
output port.}
\end{figure}
\  \  \  \

This section first introduces the phase estimation model based on MZI with
CS and PASVS as the input states. As illustrated in Fig. 1, the balanced MZI
includes two $50\colon 50$ optical beam splitters (BS1 and BS2) and a phase
shifter. Based on the research by Yurke \emph{et al.} \cite{22}, using
angular momentum operators in the Schwinger representation, the equivalent
operators for BS1 and BS2 are represented as $B_{1}=\exp \left[ -i\pi
(a^{\dagger }b+ab^{\dagger })/4\right] $ and $B_{2}=\exp \left[ i\pi
(a^{\dagger }b+ab^{\dagger })/4\right] $, respectively. These operators
satisfy the transformation relations%
\begin{equation}
B_{1}^{\dagger }\binom{a}{b}B_{1}=\frac{\sqrt{2}}{2}\left(
\begin{array}{cc}
1 & -i \\
-i & 1%
\end{array}%
\right) \binom{a}{b},  \label{1}
\end{equation}%
and%
\begin{equation}
B_{2}^{\dagger }\binom{a}{b}B_{2}=\frac{\sqrt{2}}{2}\left(
\begin{array}{cc}
1 & i \\
i & 1%
\end{array}%
\right) \binom{a}{b}.  \label{2}
\end{equation}%
The phase shifter operator is expressed as $U\left( \phi \right) =\exp \left[
i\phi \left( b^{\dagger }b\right) \right] $, and the associated
transformation relation is $U^{\dagger }\left( \phi \right) bU\left( \phi
\right) =e^{i\phi }b$.

Our scheme aims to enhance the precision of phase measurement by utilizing
CS and PASVS as input to the MZI. The input state is expressed as $%
\left
\vert \psi \right \rangle _{in}=\left \vert \alpha \right \rangle
_{a}\otimes \left \vert r,m\right \rangle _{b}$, where $\left \vert \alpha
\right \rangle _{a}$ represents the CS at the $a$-mode input, and $%
\left
\vert r,m\right \rangle _{b}$ denotes the PASVS at the $b$-mode
input. The CS satisfies $a\left \vert \alpha \right \rangle _{a}=\alpha
\left \vert \alpha \right \rangle _{a}$, where the amplitude parameter $%
\alpha =\left
\vert \alpha \right \vert e^{i\theta }$. To simplify, we set $%
\theta =0$ ($\alpha =\left \vert \alpha \right \vert $), making it more
advantageous for enhanced phase estimation \cite{38,48}. The state $%
\left
\vert r,m\right
\rangle _{b}$ is generated through non-Gaussian
operations involving $m$th-order photon addition $b^{\dagger m}$ to the SVS,
defined as%
\begin{equation}
\left \vert r,m\right \rangle _{b}=\frac{1}{\sqrt{P_{m}}}b^{\dagger
m}S\left( r\right) \left \vert 0\right \rangle _{b},  \label{3}
\end{equation}%
where $P_{m}$\ represents the normalization factor, and $S\left( r\right)
=\exp \left[ r\left( b^{2}-b^{\dag 2}\right) /2\right] $ denotes the
squeezing operator with the squeezing parameter $r$. In particular, when $%
m=0 $, this corresponds to the scenario where CS mixed with SVS is utilized
as input. For computational ease, we utilize $S\left( r\right) \left \vert
0\right \rangle _{b}=$sech$^{1/2}r\exp \left[ -b^{\dag 2}\tanh r/2\right]
\left \vert 0\right \rangle _{b}$ and $b^{\dagger m}=\frac{\partial ^{m}}{%
\partial \tau ^{m}}\left. e^{b^{\dagger }\tau }\right \vert _{\tau =0}$.
This representation leads to another expression of $\left \vert
r,m\right
\rangle _{b}$ as%
\begin{equation}
\left \vert r,m\right \rangle _{b}=\frac{\left. \frac{\partial ^{m}}{%
\partial \tau ^{m}}\exp \left[ b^{\dagger }\tau -\frac{1}{2}b^{\dagger
2}\tanh r\right] \right \vert _{\tau =0}\left \vert 0\right \rangle _{b}}{%
\sqrt{P_{m}\cosh r}},  \label{4}
\end{equation}%
thus, $P_{m}$ and $\bar{n}_{b}$ can be derived as%
\begin{eqnarray}
P_{m} &=&\left. \frac{\partial ^{2m}}{\partial t^{m}\partial \tau ^{m}}e^{-%
\frac{\sinh 2r}{4}\left( t^{2}+\tau ^{2}\right) }e^{t\tau \cosh ^{2}r}\right
\vert _{t=\tau =0},  \label{5} \\
\bar{n}_{b} &=&\left. _{b}\left \langle r,m\right \vert b^{\dagger }b\left
\vert r,m\right \rangle _{b}\right. =\frac{P_{m+1}}{P_{m}}-1.  \label{6}
\end{eqnarray}%
In our scheme, the total average photon number $\bar{N}=\left.
_{in}\left
\langle \psi \right \vert \left( a^{\dag }a+b^{\dagger }b\right)
\left \vert \psi \right \rangle _{in}\right. =\bar{n}_{a}+\bar{n}_{b}$ with $%
\bar{n}_{a}=\alpha ^{2}$.

Several studies have demonstrated that for specific path-symmetric state
inputs to the MZI, parity detection using photon-number-resolving detectors
can achieve the phase sensitivity saturating the QCRB. In this case, parity
detection constitutes an optimal measurement scheme, leading to significant
theoretical and experimental advancements \cite%
{32,33,34,35,36,37,38,45,47,48,49,50,51,52,60}. In the model illustrated in
Fig. 1, parity detection is employed to measure the phase shift at the $b$%
-mode output port of the MZI, where the parity operator is defined as $\Pi
_{b}=\left( -1\right) ^{b^{\dagger }b}=e^{i\pi b^{\dagger }b}$.

To investigate the actual measurements, we simulated photon loss in the $b$%
-mode of the MZI by introducing a virtual beam splitter between BS1 and the
phase shifter. The corresponding transformation relation is

\begin{equation}
B_{f}^{\dagger }\left(
\begin{array}{c}
b \\
b_{f}%
\end{array}%
\right) B_{f}=\left(
\begin{array}{cc}
\sqrt{1-l} & \sqrt{l} \\
-\sqrt{l} & \sqrt{1-l}%
\end{array}%
\right) \left(
\begin{array}{c}
b \\
b_{f}%
\end{array}%
\right) ,  \label{7}
\end{equation}%
where $b_{f}$\ is the photon annihilation operator of the dissipative mode
in which the vacuum noise $\left \vert 0\right \rangle _{f}$ is located, and
$B_{f}$ represents the equivalent operator of a virtual beam splitter with
reflectivity corresponding to a photon loss rate $l$. Specifically, $l=0$
and $l=1$ correspond to a lossless scenario and complete absorption,
respectively.

To simplify the study of the parity detection scheme, we define an
equivalent operator for parity detection that encompasses the entire lossy
MZI, denoted as $\Pi _{b}^{loss}$, as follows:%
\begin{equation}
\Pi _{b}^{loss}=\left. _{f}\left \langle 0\right \vert B_{1}^{\dagger
}B_{f}^{\dagger }U^{\dagger }\left( \phi \right) B_{2}^{\dagger }\Pi
_{b}B_{2}U\left( \phi \right) B_{f}B_{1}\left \vert 0\right \rangle
_{f}\right. .  \label{8}
\end{equation}%
Using the normal ordering form of $\Pi _{b}^{loss}$, it is convenient to
compute its average value with respect to the input state, i.e., $%
\left
\langle \Pi _{b}\right \rangle =\left. _{in}\left \langle \psi
\right
\vert \Pi _{b}^{loss}\left \vert \psi \right \rangle _{in}\right. $.
The value $\left \langle \Pi _{b}\right \rangle $ characterizes the signals
of parity detection. The derivation procedure and specific expressions for
the normal ordering form of $\Pi _{b}^{loss}$, as well as its average value $%
\left
\langle \Pi _{b}\right \rangle $, are provided in Appendix A.
\begin{figure}[tbh]
\label{Fig2} \centering
\subfigure{
\begin{minipage}[b]{0.5\textwidth}
\includegraphics[width=0.83\textwidth]{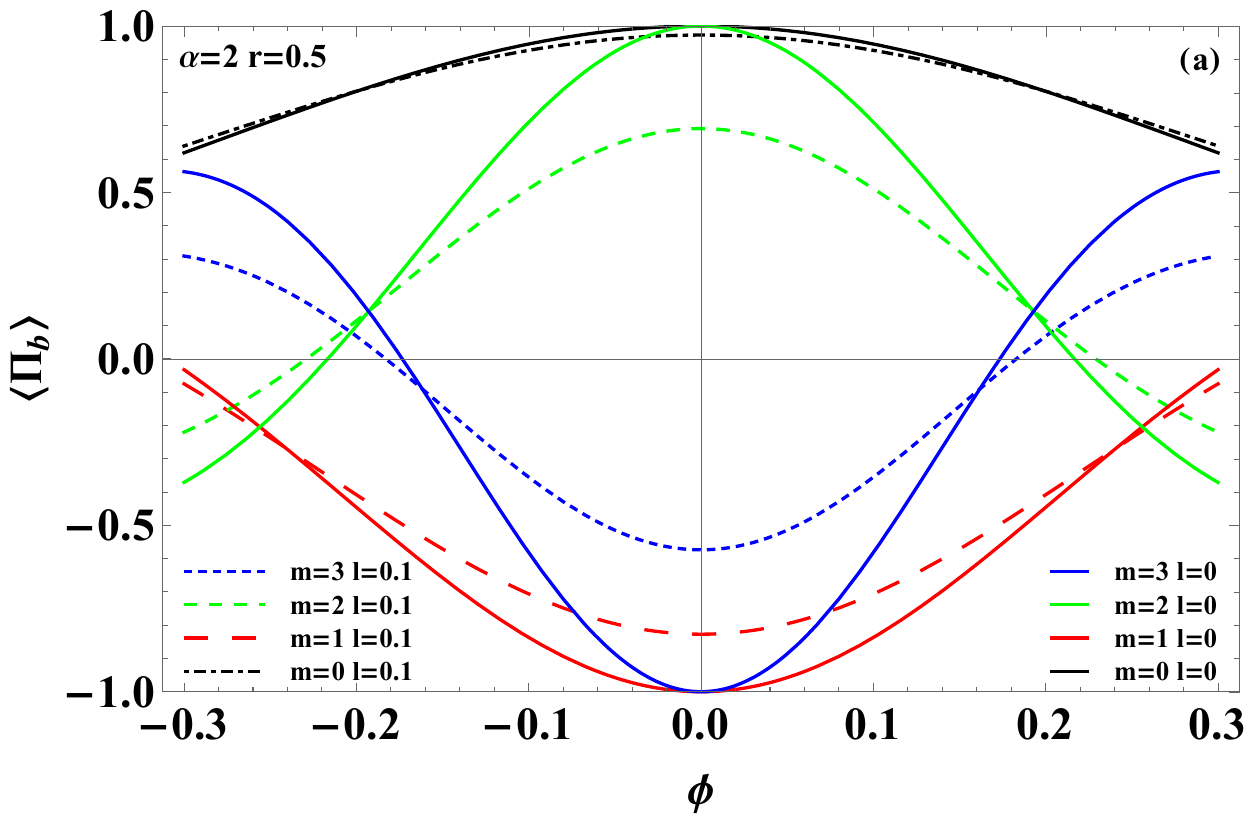}\\
\includegraphics[width=0.83\textwidth]{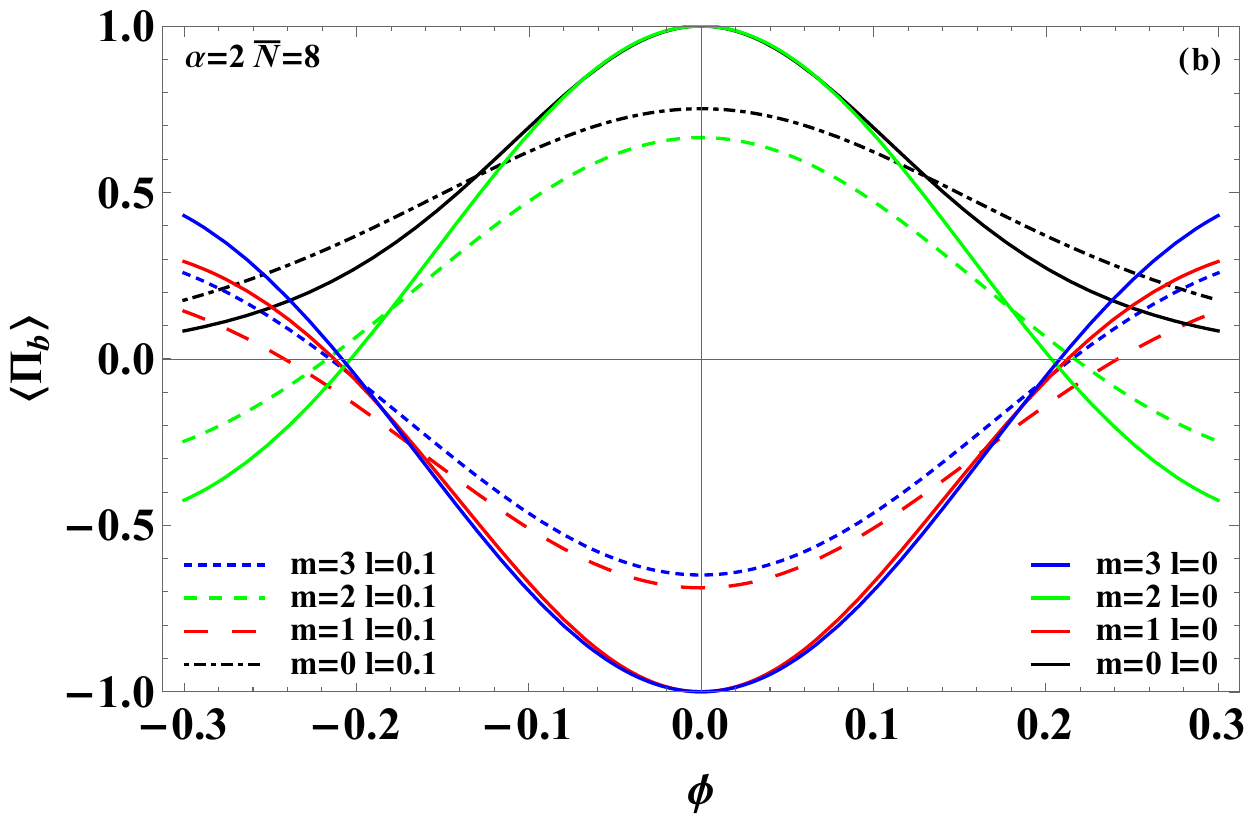}
\end{minipage}}
\caption{The variation of the parity signal $\langle \Pi _{b}\rangle $ with
phase shift $\protect \phi $ for the ideal case ($l=0$) and the photon loss
case ($l=0.1$), the addition photon number $m=0,1,2,3$ and the coherent
amplitude $\protect \alpha =2$, (a) with the fixed squeezing parameter $r=0.5$%
, and (b) with the fixed total average photon number $\bar{N}=8$.}
\end{figure}

Using Eq. (A11) for the average value $\left \langle \Pi _{b}\right \rangle $%
, we can graph its behavior as a function of the phase shift $\phi $. Fig. 2
illustrates $\left \langle \Pi _{b}\right \rangle $ with $\phi $ for both
the ideal ($l=0$) and photon loss ($l=0.1$) cases. In Fig. 2(a), with fixed $%
\alpha =2$ and $r=0.5$, it is clearly demonstrated that the central peak or
trough of $\left \langle \Pi _{b}\right \rangle $ at $\phi =0$ narrows as
the addition photon number $m$ increases. This indicates that increasing $m$
can effectively enhance the phase resolution of the parity signal.\ In
addition, as shown in Fig. 2(b), when $\alpha =2$ and $\bar{N}=8$, the
distribution of the central peak or trough remains largely consistent across
different $m$ due to the energy constraint imposed by the fixed the total
average photon number $\bar{N}$. Furthermore, as illustrated in Fig. 2,
photon loss results in a broader central peak or trough of $\left \langle
\Pi _{b}\right \rangle $ relative to the ideal scenario. This shows that
photon loss degrades the phase resolution. Based on the comparison of the
dashed lines representing photon loss in Fig. 2(a), it is evident that the
central peak or trough narrows as $m$ increases. Thus, even in the presence
of photon loss, phase resolution can be enhanced by photon addition
operations, which implies that photon addition may improve the phase
sensitivity of parity detection.

\section{Phase sensitivity with parity detection}

For our scheme, the classical Fisher information (CFI) $F_{C}$ for the phase
shift $\phi $ can be obtained by using parity detection at the output port
of the $b$-mode for even or odd photon numbers, as follows \cite{49}:%
\begin{equation}
F_{C}=\frac{1}{P_{e}}\left( \frac{\partial P_{e}}{\partial \phi }\right)
^{2}+\frac{1}{P_{o}}\left( \frac{\partial P_{o}}{\partial \phi }\right) ^{2},
\label{9}
\end{equation}%
where $P_{e}=\left( 1+\left \langle \Pi _{b}\right \rangle \right) /2$ and $%
P_{o}=\left( 1-\left \langle \Pi _{b}\right \rangle \right) /2$ represent
the probabilities of even and odd counts, respectively. In this section, we
investigate the phase sensitivity of parity detection, also referred to as
phase uncertainty, under both ideal condition and photon loss. According to
the Cram\'{e}r-Rao bound, the error propagation formula for phase
sensitivity can be derived using Eq. (\ref{9}), i.e.,%
\begin{equation}
\Delta \phi =\frac{1}{\sqrt{F_{C}}}=\frac{\sqrt{1-\langle \Pi _{b}\rangle
^{2}}}{\left \vert \partial \left \langle \Pi _{b}\right \rangle /\partial
\phi \right \vert }.  \label{10}
\end{equation}%
By substituting Eq. (A11) into Eq. (\ref{10}), one can derive the phase
sensitivity $\Delta \phi $.

\begin{figure}[tbh]
\label{Fig3} \centering%
\subfigure{
\begin{minipage}[b]{0.5\textwidth}
\includegraphics[width=0.83\textwidth]{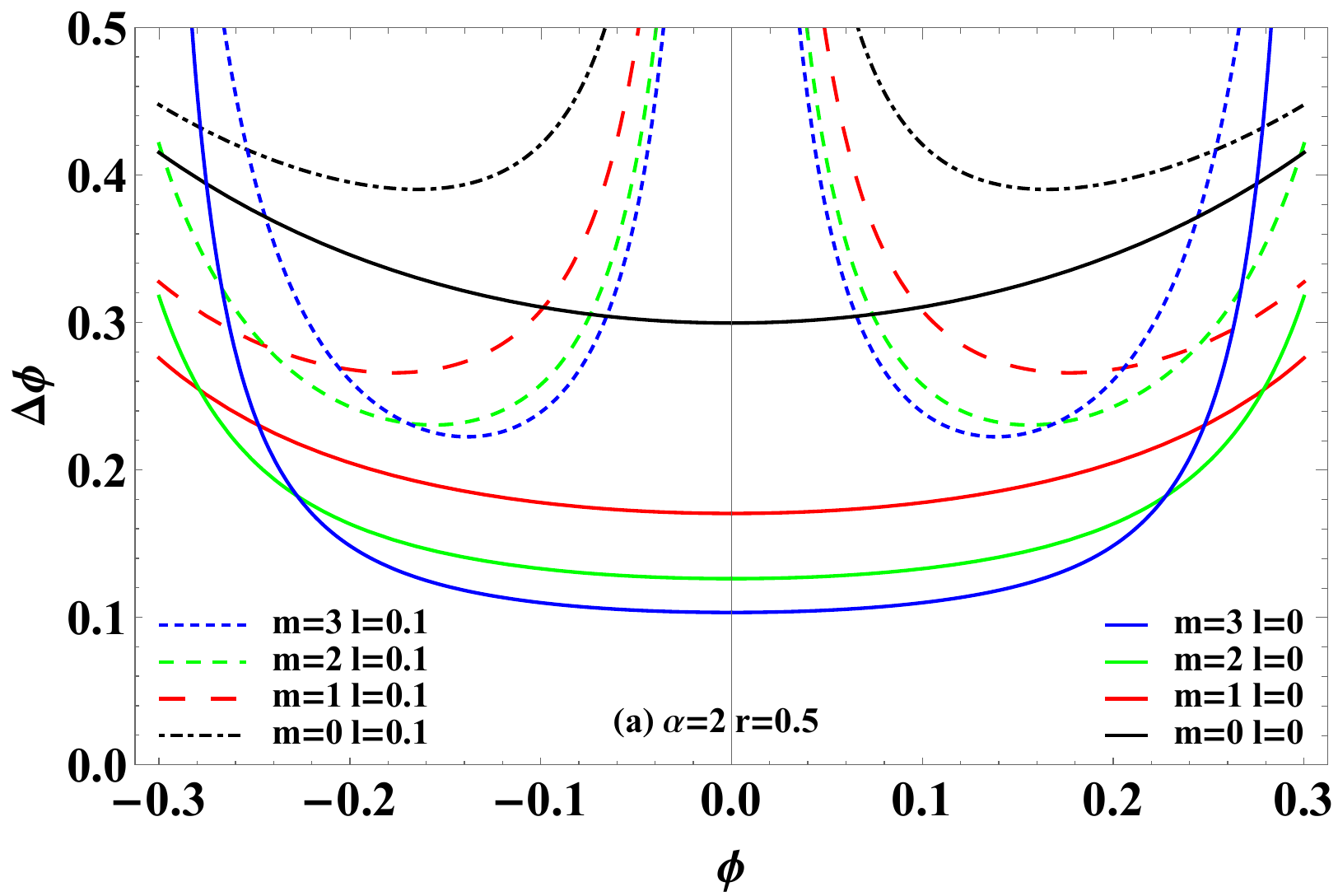}\\
\includegraphics[width=0.83\textwidth]{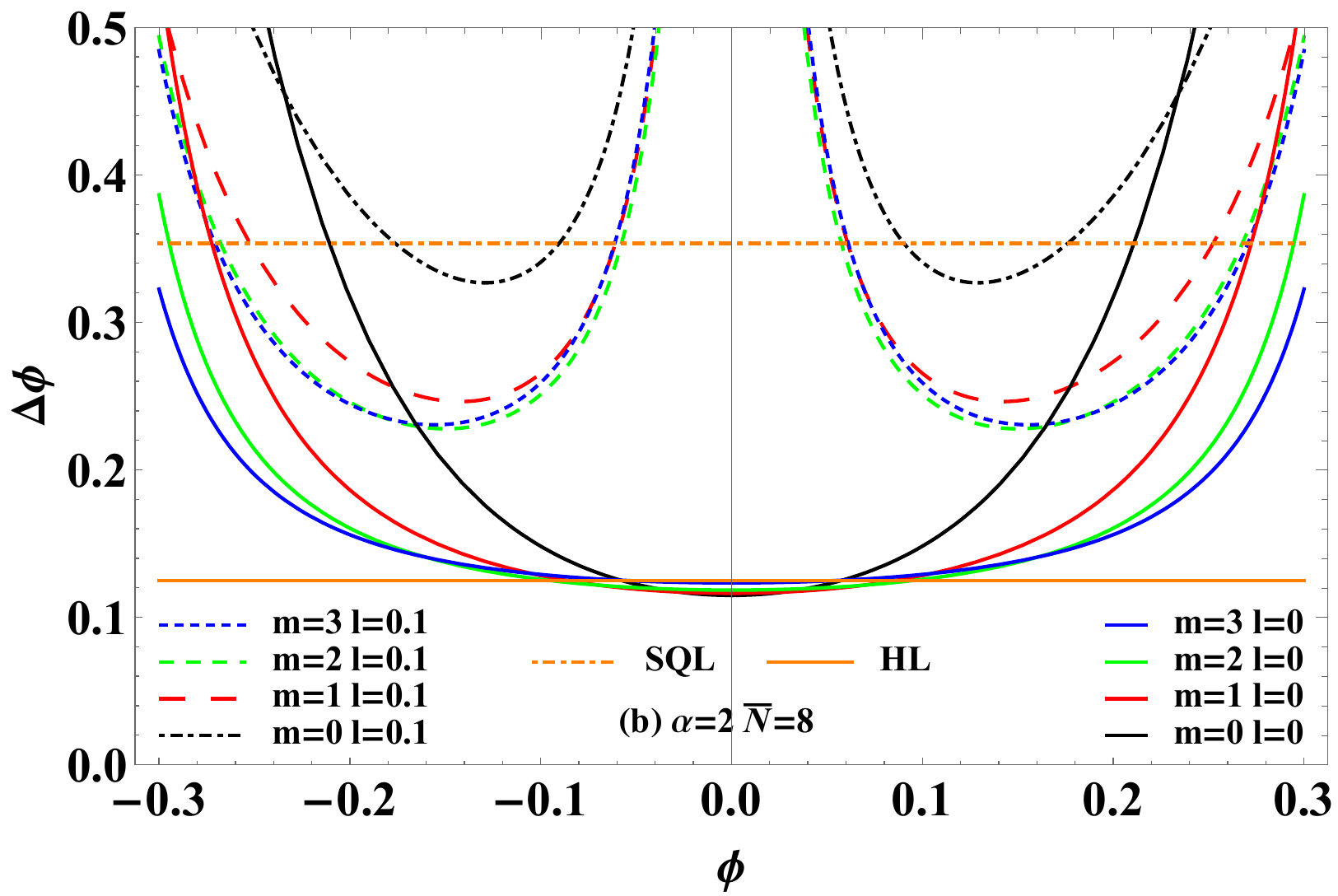}
\end{minipage}}
\caption{For the addition photon number of $m=0,1,2,3$ and the coherent
amplitude $\protect \alpha =2$, the phase sensitivity $\Delta \protect \phi $
as a function of the phase shift $\protect \phi $ for the ideal case ($l=0$)
and the photon loss case ($l=0.1$), (a) with the fixed squeezing parameter $%
r=0.5$, and (b) with the fixed total average photon number $\bar{N}=8$ and
compared with the SQL and the HL.}
\end{figure}

Fig. 3 illustrates the variation of the phase sensitivity $\Delta \phi $
with the phase shift $\phi $ for both the ideal ($l=0$) and photon loss ($%
l=0.1$) cases. As shown in the Fig. 3, in the ideal case, $\Delta \phi $
reaches its optimum when $\phi =0$. Compared with the ideal scenario, photon
loss degrades $\Delta \phi $ and causes the optimal point to deviate from $%
\phi =0$. As depicted in Fig. 3(a), when $\alpha =2$ and $r=0.5$ are given, $%
\Delta \phi $ can be effectively enhanced by increasing the addition photon
number $m$ within a certain range near the optimal point of $\phi $.
Furthermore, according to Refs. \cite{38,48,53,54}, $\Delta \phi $ yields
favorable results when $\bar{n}_{a}=\bar{n}_{b}$ and $\bar{N}$ is fixed at a
suitably large value. Thus,\ to further investigate the enhancement effect
of our scheme on phase sensitivity $\Delta \phi $, Fig. 3(b) illustrates $%
\Delta \phi $ as a function of $\phi $ for $\alpha =2$ and $\bar{N}=8$ ($%
\bar{n}_{a}=\bar{n}_{b}=4$), and compares it with the SQL and the HL. It is
shown that the ideal $\Delta \phi $ significantly surpasses the SQL and
approaches the HL near the optimal point $\phi =0$. Moreover, near $\phi =0$%
, $\Delta \phi $ remains essentially the same for $m=0,1,2,3$ due to the
energy constraint imposed by the fixed $\bar{N}$, while $\Delta \phi $
improves with increasing $m$ when $\phi $ deviates from $0$ for a certain
range. Compared with the ideal case, the photon loss with $l=0.1$ makes the
phase sensitivity $\Delta \phi $ deteriorate, and the optimal point deviates
from $\phi =0$. However, by using the photon addition operations of $m=1,2,3$%
, $\Delta \phi $ can be significantly improved relative to $m=0$, and $%
\Delta \phi $ can still effectively break through the SQL within a certain
range.
\begin{figure}[tbh]
\label{Fig4} \centering
\subfigure{
\begin{minipage}[b]{0.5\textwidth}
\includegraphics[width=0.83\textwidth]{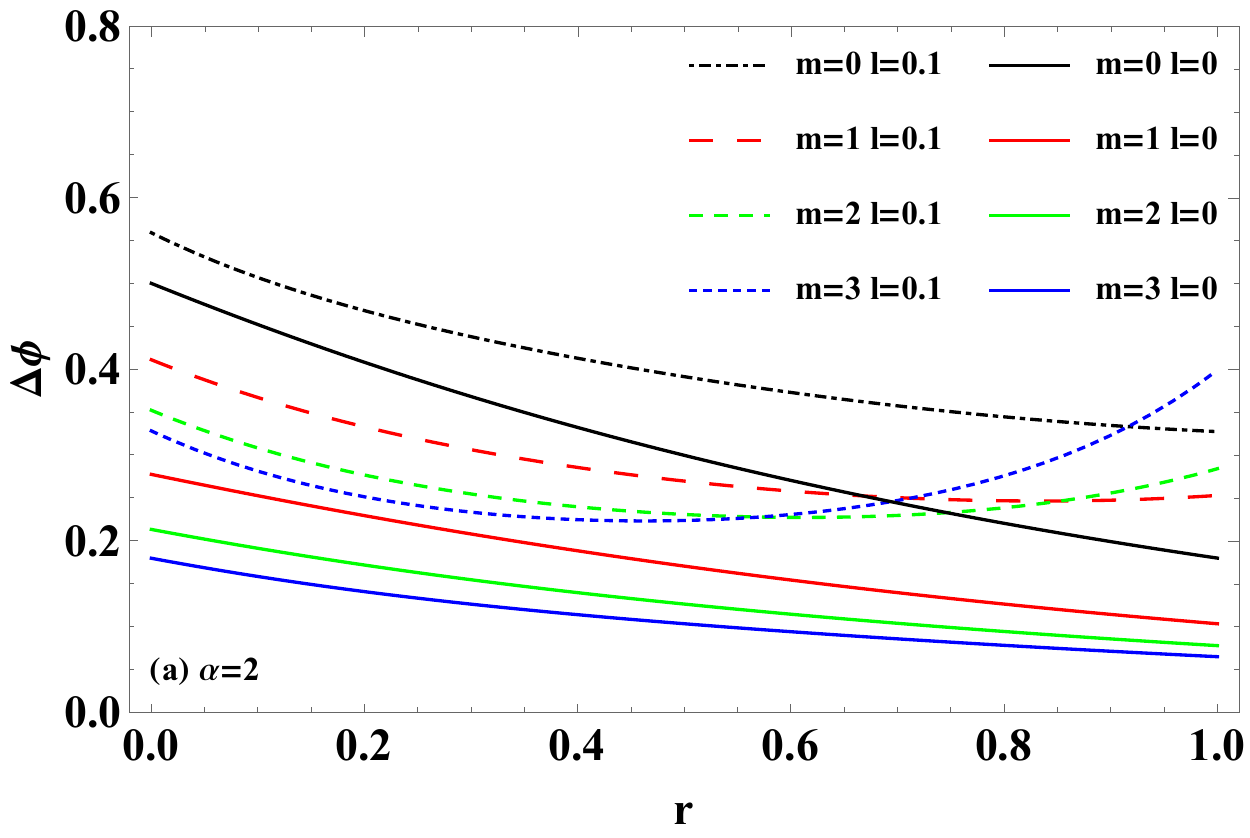}\\
\includegraphics[width=0.83\textwidth]{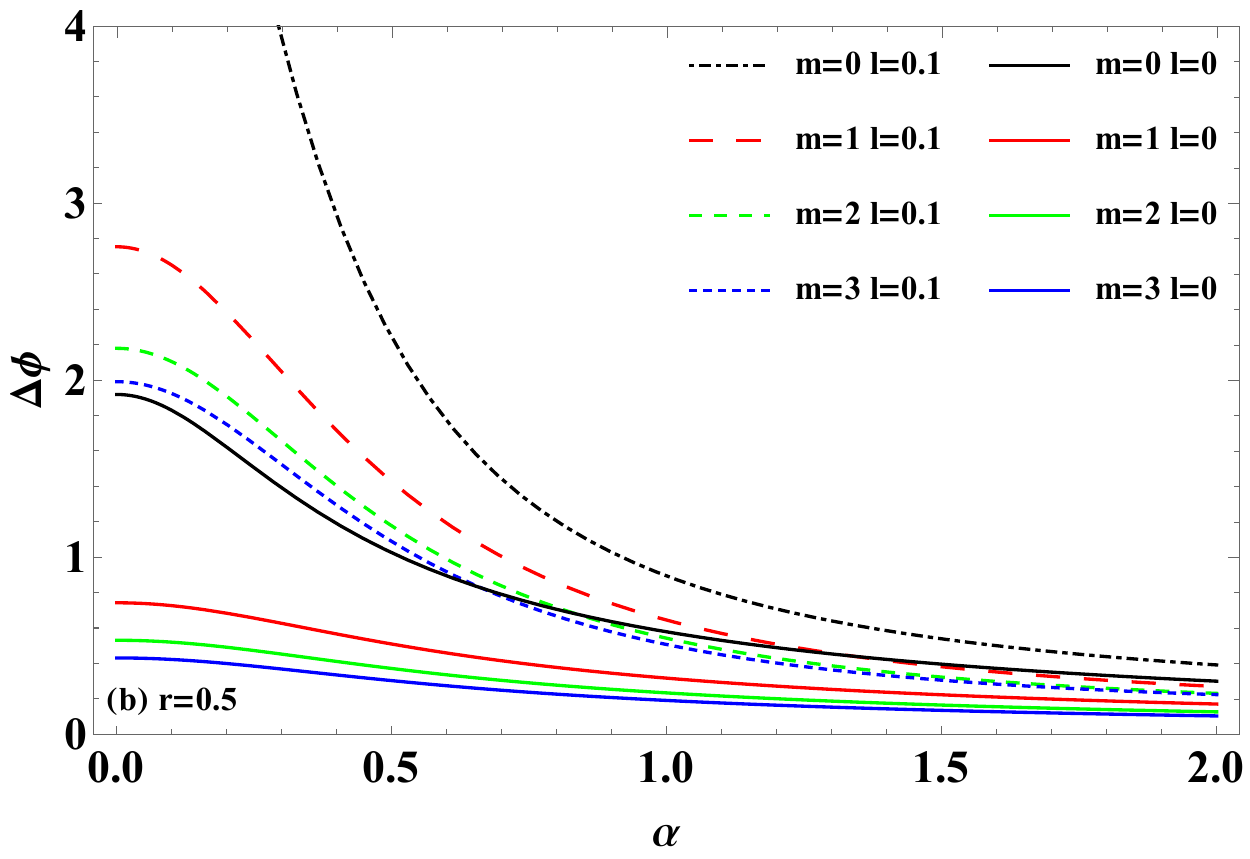}
\end{minipage}}
\caption{In the ideal case ($l=0$) with fixed the phase shift $\protect \phi %
=10^{-4}$ and the photon loss case ($l=0.1$) with fixed $\protect \phi =0.15$%
, for the addition photon number $m=0,1,2,3$, (a) the phase sensitivity $%
\Delta \protect \phi $ as a function of the squeezing parameter $r$ with the
coherent amplitude $\protect \alpha =2$, (b) $\Delta \protect \phi $ as a
function of $\protect \alpha $ with $r=0.5$.}
\end{figure}

As shown in Fig. 4, the phase sensitivity $\Delta \phi $ as a function of
the squeezing parameter $r$ and coherent amplitude $\alpha $ is depicted
under both the ideal ($l=0$) and photon loss ($l=0.1$) scenarios. It is
evident that the ideal $\Delta \phi $ can be enhanced by increasing $r$, $%
\alpha $ and the addition photon number $m$ for a small phase shift $\phi
=10^{-4}$. Despite the impact of photon loss, when an appropriate $\phi
=0.15 $ is selected (as illustrated in Fig. 3, photon loss results in the
optimal $\phi $ point\ deviating from $0$), $\Delta \phi $ can still be
significantly enhanced by increasing $r$, $\alpha $ and $m$ within a certain
range. Specifically, in Fig. 4(a), it is observed that under the photon
loss, as $m$ increases, the range over which $\Delta \phi $ improves with
increasing $r$ is confined to relatively small values of $r$. Actually, the
smaller squeezing parameter $r$ values are more experimentally feasible.
\begin{figure}[tbh]
\label{Fig5} \centering%
\subfigure{
\begin{minipage}[b]{0.5\textwidth}
\includegraphics[width=0.83\textwidth]{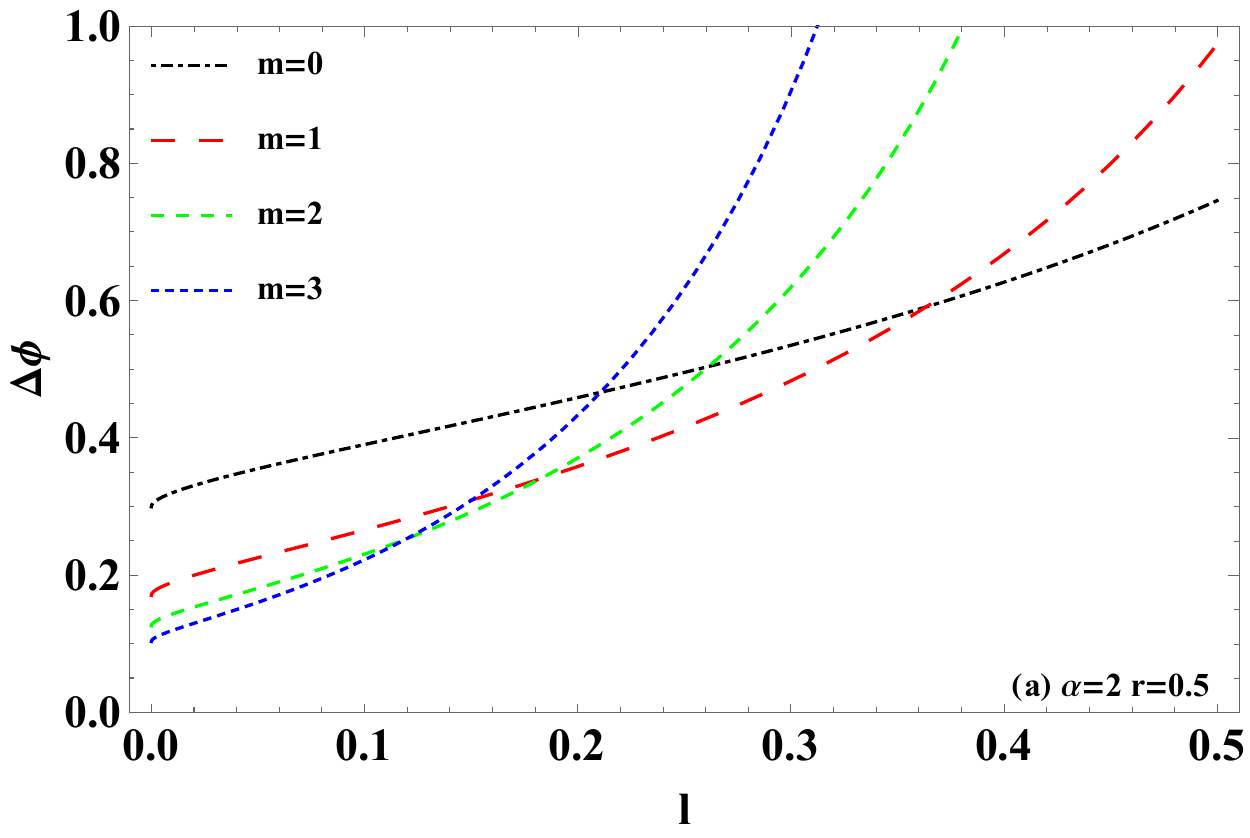}\\
\includegraphics[width=0.83\textwidth]{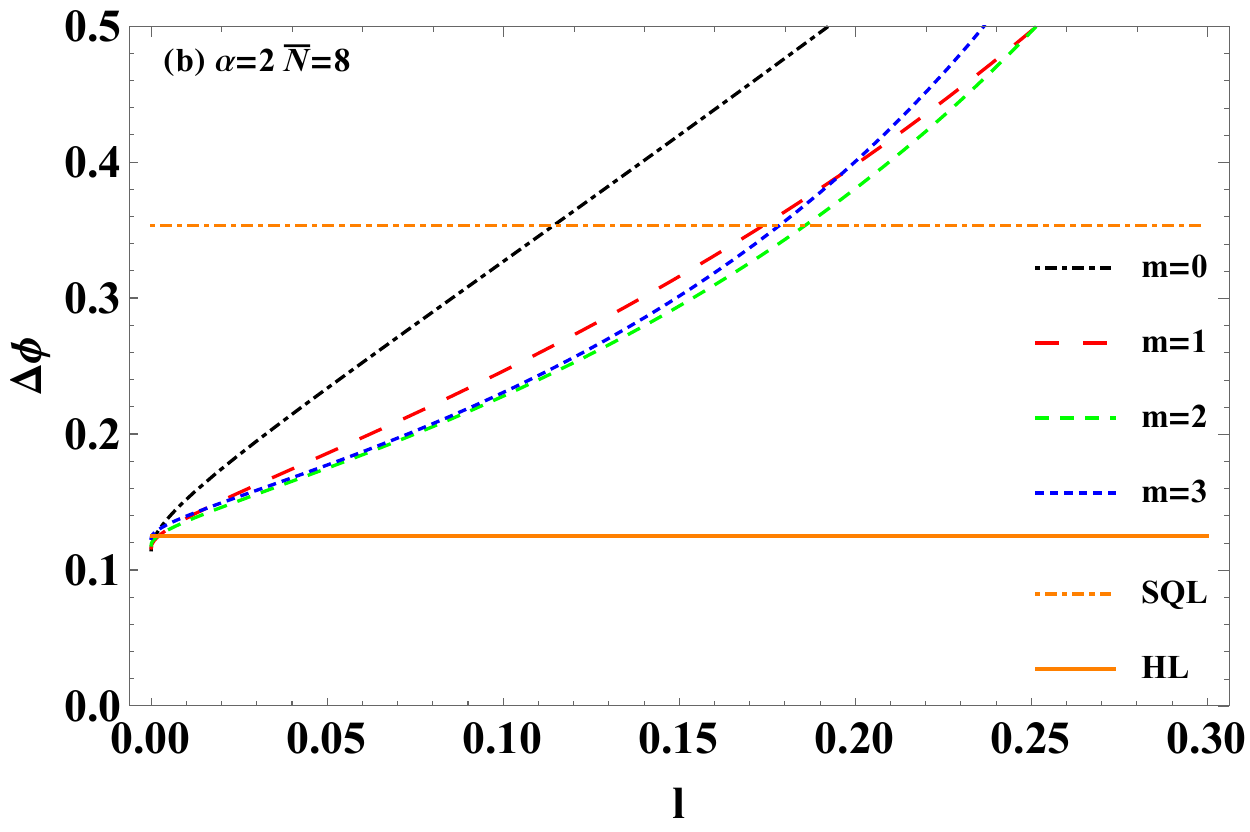}
\end{minipage}}
\caption{For the addition photon number $m=0,1,2,3$, the coherent amplitude $%
\protect \alpha =2$, and optimized the phase shift $\protect \phi $, the phase
sensitivity $\Delta \protect \phi $ as a function of loss rate $l$ (a) for
given the squeezing parameter $r=0.5$, and (b) for given the total average
photon number $\bar{N}=8$, and compared with the SQL and the HL.}
\end{figure}

To further analyze the influence of photon loss on phase sensitivity $\Delta
\phi $, as shown in Fig. 5, we present the variation of $\Delta \phi $ with
respect to the loss rate $l$ (phase shift $\phi $ is optimized). Although $%
\Delta \phi $ deteriorates with the increase of the loss rate $l$ (see Fig.
5(a)), when $\alpha =2$ and $r=0.5$ are fixed, $\Delta \phi $ can still be
significantly improved by the photon addition operations of $m=1,2,3$ within
a certain range of relatively small $l$. Fig. 5(b) shows $\Delta \phi $
versus $l$ for $\alpha =2$ and $\bar{N}=8$, comparing the SQL and HL. It can
be found that $\Delta \phi $ can reach the HL when $l$ is very small, and in
the case of photon loss, $\Delta \phi $ can still effectively break through
the SQL in a certain range of $l$. Compared to the case of $m=0$, $\Delta
\phi $ of $m=1,2,3$ has a significant improvement and can exceed the SQL in
a larger range of $l$. This indicates that the photon addition operations
enhance the robustness of the measured system.

\section{The QFI in MZI}

The QFI quantifies the maximum amount of obtainable information regarding
the phase shift $\phi $ of a quantum system, independent of any detection
scheme, and it is an upper bound of the CFI. The optimal error bound for
phase sensitivity, known as the QCRB, is given by \cite{55,56,57}
\begin{equation}
\Delta \phi _{QCRB}=\frac{1}{\sqrt{F_{Q}}},  \label{11}
\end{equation}%
where $F_{Q}$ is the QFI. For a pure state, the QFI under ideal condition
can be calculated as \cite{58}

\begin{equation}
F_{Q}=4\left[ \left \langle \psi _{\phi }^{\prime }|\psi _{\phi }^{\prime
}\right \rangle -\left \vert \left \langle \psi _{\phi }^{\prime }|\psi
_{\phi }\right \rangle \right \vert ^{2}\right] ,  \label{12}
\end{equation}%
where $\left \vert \psi _{\phi }\right \rangle =U\left( \phi \right)
B_{1}\left \vert \psi \right \rangle _{in}$ is the quantum state before BS2
in the lossless MZI, and $\left \vert \psi _{\phi }^{^{\prime
}}\right
\rangle =\partial \left \vert \psi _{\phi }\right \rangle
/\partial \phi $. Thus, the QFI can be further simplified as%
\begin{equation}
F_{Q}=4\left \langle \Delta ^{2}n_{b}\right \rangle =4\left[ \left \langle
n_{b}^{2}\right \rangle -\left \langle n_{b}\right \rangle ^{2}\right] ,
\label{13}
\end{equation}%
where $n_{b}=b^{\dagger }b$, and $\left \langle \cdot \right \rangle =\left.
\left \langle \Psi _{S}\right \vert \cdot \left \vert \Psi
_{S}\right
\rangle \right. $, with $\left \vert \Psi _{S}\right \rangle
=B_{1}\left
\vert \psi \right \rangle _{in}$.

\begin{figure}[tbh]
\label{Fig6} \centering \includegraphics[width=0.83\columnwidth]{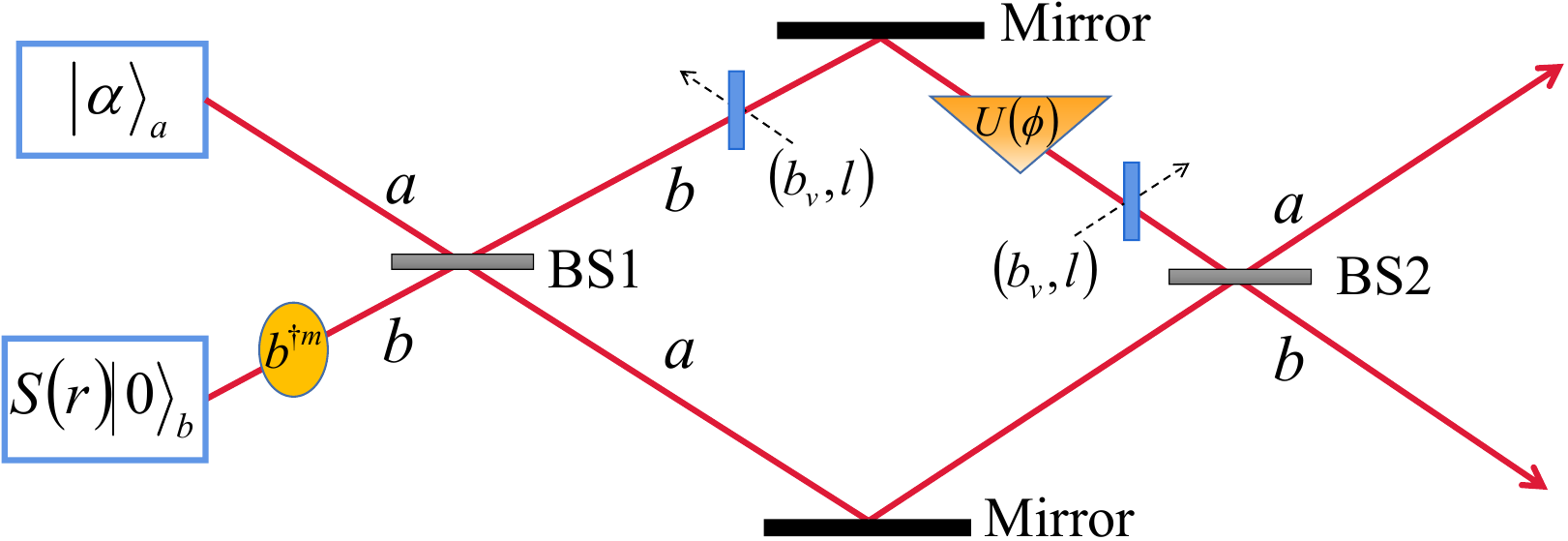}
\caption{Theoretical model of lossy MZI using CS mixed with PASVS as input.
Two virtual optical beam splitters, positioned before and after the phase
shifter, simulate photon loss. The parameter $l$ represents the
reflectivity, or loss rate, of the optical beam splitter, while $b_{v}$
denotes the vacuum operator.}
\end{figure}

For the initial pure state $\left \vert \Psi _{S}\right \rangle $ in a probe
system $S$ with a lossy MZI, we introduce the orthogonal states $\left \vert
j_{E}\right \rangle $ of the environment $E$ and the Kraus operator $\hat{\Pi%
}_{j}\left( \phi \right) $ to characterize the behavior of $\left \vert \Psi
_{S}\right \rangle $ as it passes through the phase shifter, accounting for
photon loss. A diagram illustrating photon loss inside the MZI is provided
in Fig. 6. According to Escher \emph{et al.}'s methods for calculating QFI
in open quantum systems \cite{59}, the quantum states of $\left \vert \Psi
_{S}\right \rangle $ and the vacuum noise in the environment $\left \vert
0_{E}\right \rangle $, after undergoing unitary evolution $U_{S+E}\left(
\phi \right) $ that accounts for photon loss, can be expressed in the
extended space of $S+E$ as

\begin{eqnarray}
\left \vert \Psi _{S+E}\right \rangle &=&U_{S+E}\left( \phi \right) \left
\vert \Psi _{S}\right \rangle \left \vert 0_{E}\right \rangle  \notag \\
&=&\sum_{j=0}^{\infty }\hat{\Pi}_{j}\left( \phi \right) \left \vert \Psi
_{S}\right \rangle \left \vert j_{E}\right \rangle .  \label{14}
\end{eqnarray}%
Indeed, after photon loss, the quantum state $\left \vert \Psi
_{S}\right
\rangle $ transitions into a mixed state; however, we employ
this approach to treat the quantum state $\left \vert \Psi
_{S+E}\right
\rangle $ in $S+E$ as a pure state. In this context, for the
entire purified system, the QFI under photon loss can be expressed as%
\begin{equation}
F_{Q}\leqslant C_{Q}\left[ \left \vert \Psi _{S}\right \rangle ,\hat{\Pi}%
_{j}\left( \phi \right) \right] =4\left[ \left \langle \hat{H}_{1}\right
\rangle -\left \vert \left \langle \hat{H}_{2}\right \rangle \right \vert
^{2}\right] ,  \label{15}
\end{equation}%
where the lower bound of $C_{Q}\left[ \left \vert \Psi _{S}\right \rangle ,%
\hat{\Pi}_{j}\left( \phi \right) \right] $ is demonstrated to be the QFI for
a reduced system \cite{59}, and $\hat{H}_{1,2}$ are Hermitian operators
defined as%
\begin{eqnarray}
\hat{H}_{1} &=&\sum \limits_{j=0}^{\infty }\frac{d\hat{\Pi}_{j}^{\dagger
}\left( \phi \right) }{d\phi }\frac{d\hat{\Pi}_{j}\left( \phi \right) }{%
d\phi },  \label{16} \\
\hat{H}_{2} &=&i\sum \limits_{j=0}^{\infty }\frac{d\hat{\Pi}_{j}^{\dagger
}\left( \phi \right) }{d\phi }\hat{\Pi}_{j}\left( \phi \right) ,  \label{17}
\end{eqnarray}%
where $\hat{\Pi}_{j}\left( \phi \right) $ is the Kraus operator, i.e.,

\begin{equation}
\hat{\Pi}_{j}\left( \phi \right) =\sqrt{\frac{l^{j}}{j!}}e^{i\phi \left(
b^{\dagger }b-\gamma j\right) }\left( 1-l\right) ^{\frac{b^{\dagger }b}{2}%
}b^{j},  \label{18}
\end{equation}%
where $\gamma =0$ or $-1$ correspond to photon loss before or after the
linear phase shifter, respectively. By optimizing $\gamma $, we can derive $%
C_{Q\min }$. Thus, the QFI in the presence of photon loss is obtained as
\cite{59}

\begin{equation}
F_{Q}=\frac{4\left( 1-l\right) \left \langle n_{b}\right \rangle \left
\langle \Delta ^{2}n_{b}\right \rangle }{l\left \langle \Delta
^{2}n_{b}\right \rangle +\left( 1-l\right) \left \langle n_{b}\right \rangle
},  \label{19}
\end{equation}%
where, $\left \langle n_{b}\right \rangle $ and $\left \langle
n_{b}^{2}\right \rangle $ can be obtained by setting $w$ of $\left \langle
n_{b}^{w}\right \rangle $ in Eq. (B3) to $1$ and $2$, respectively (see
Appendix B for details).
\begin{figure}[tbh]
\label{Fig7} \centering%
\subfigure{
\begin{minipage}[b]{0.5\textwidth}
\includegraphics[width=0.83\textwidth]{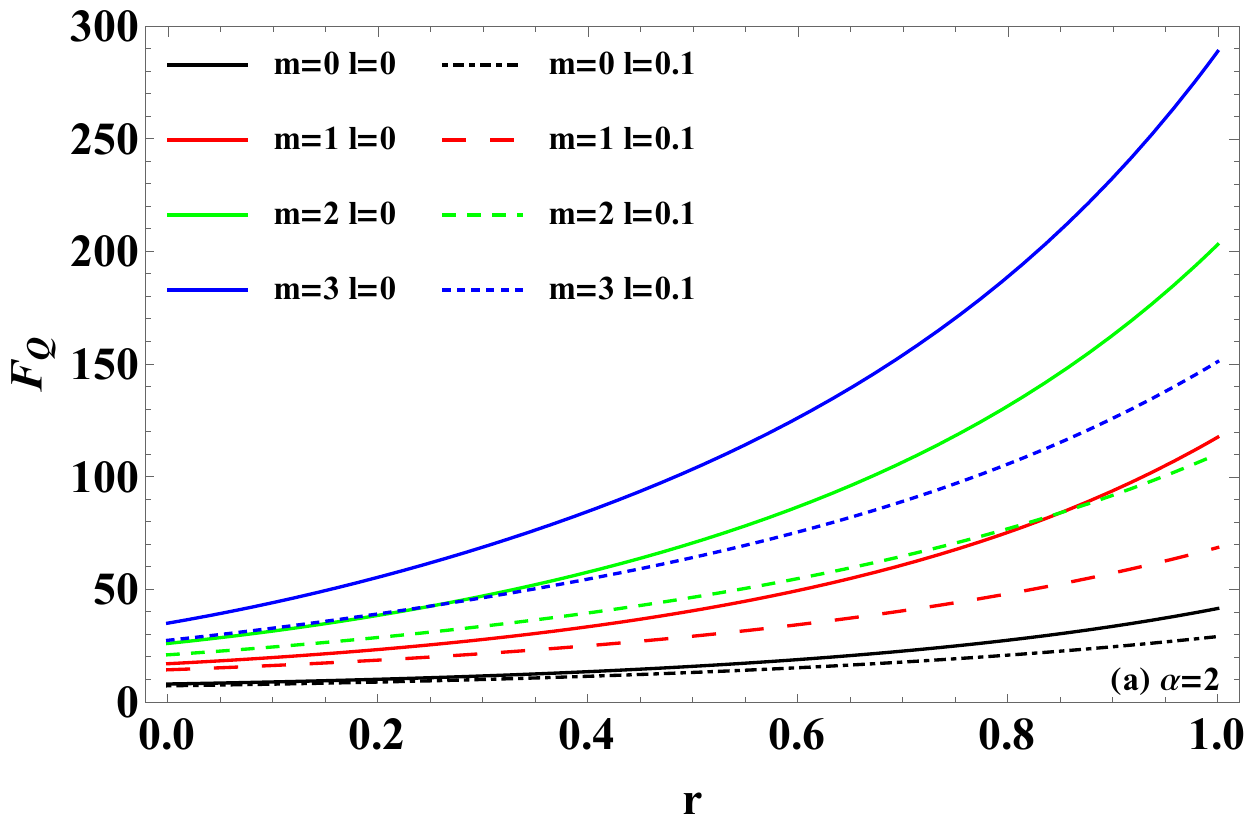}\\
\includegraphics[width=0.83\textwidth]{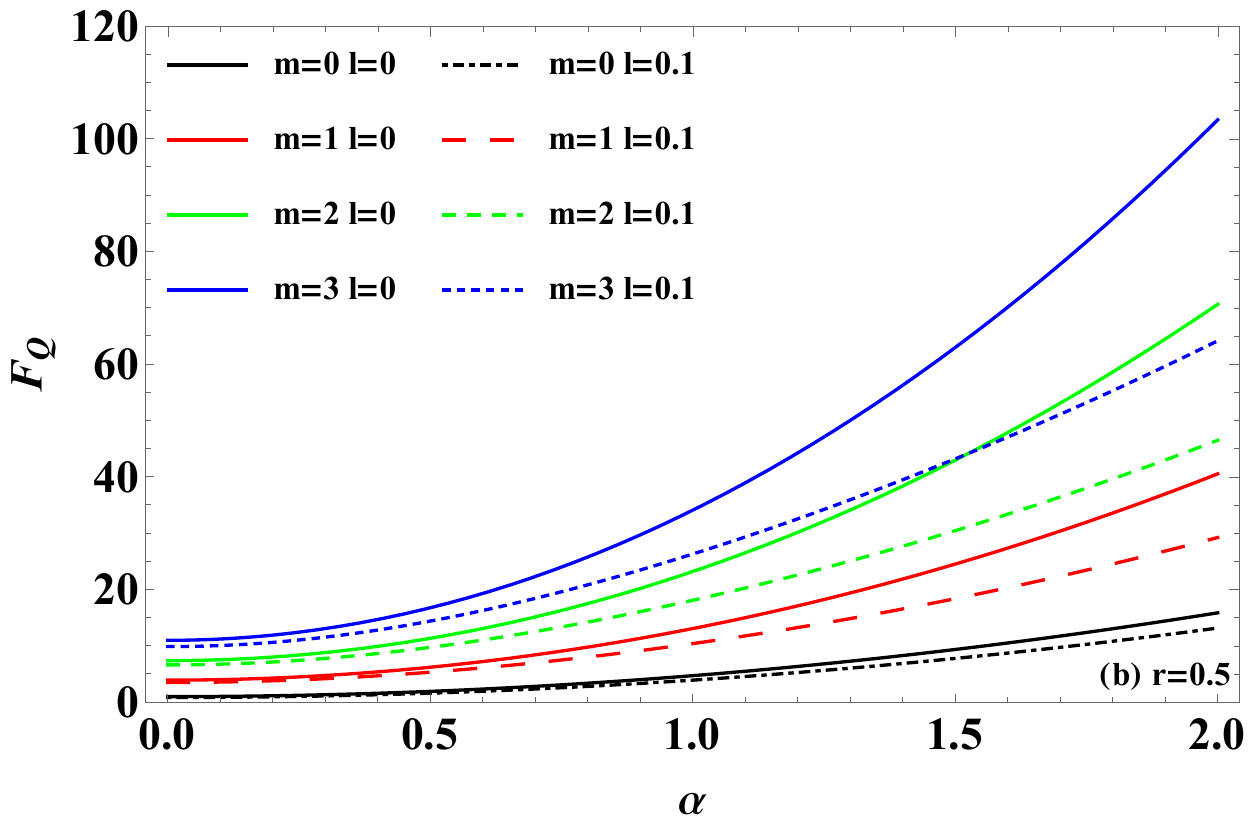}
\end{minipage}}
\caption{For the addition photon number $m=0,1,2,3$, in the ideal case of $%
l=0$, and in the photon loss case of $l=0.1$, (a) the variation of QFI $%
F_{Q} $ with the squeezing parameter $r$ for given the coherent amplitude $%
\protect \alpha =2$, and (b) the variation of $F_{Q}$ with $\protect \alpha $
for given $r=0.5$.}
\end{figure}

Fig. 7 illustrates the variation of the QFI with respect to the input state
parameters and the impact of photon loss ($l=0.1$) on the QFI. It is clear
that, although photon loss decreases the QFI $F_{Q}$, increasing the
squeezing parameter $r$, the coherent amplitude $\alpha $, and the addition
photon number $m$ can effectively increase $F_{Q}$, thereby improving the
QCRB as indicated by Eq. (\ref{11}) ($\Delta \phi _{QCRB}=1/\sqrt{F_{Q}}$),
in both the ideal and photon loss cases. Thus, the proposed input scheme can
effectively improve the accuracy of phase measurement.
\begin{figure}[tbh]
\label{Fig8} \centering%
\subfigure{
\begin{minipage}[b]{0.5\textwidth}
\includegraphics[width=0.83\textwidth]{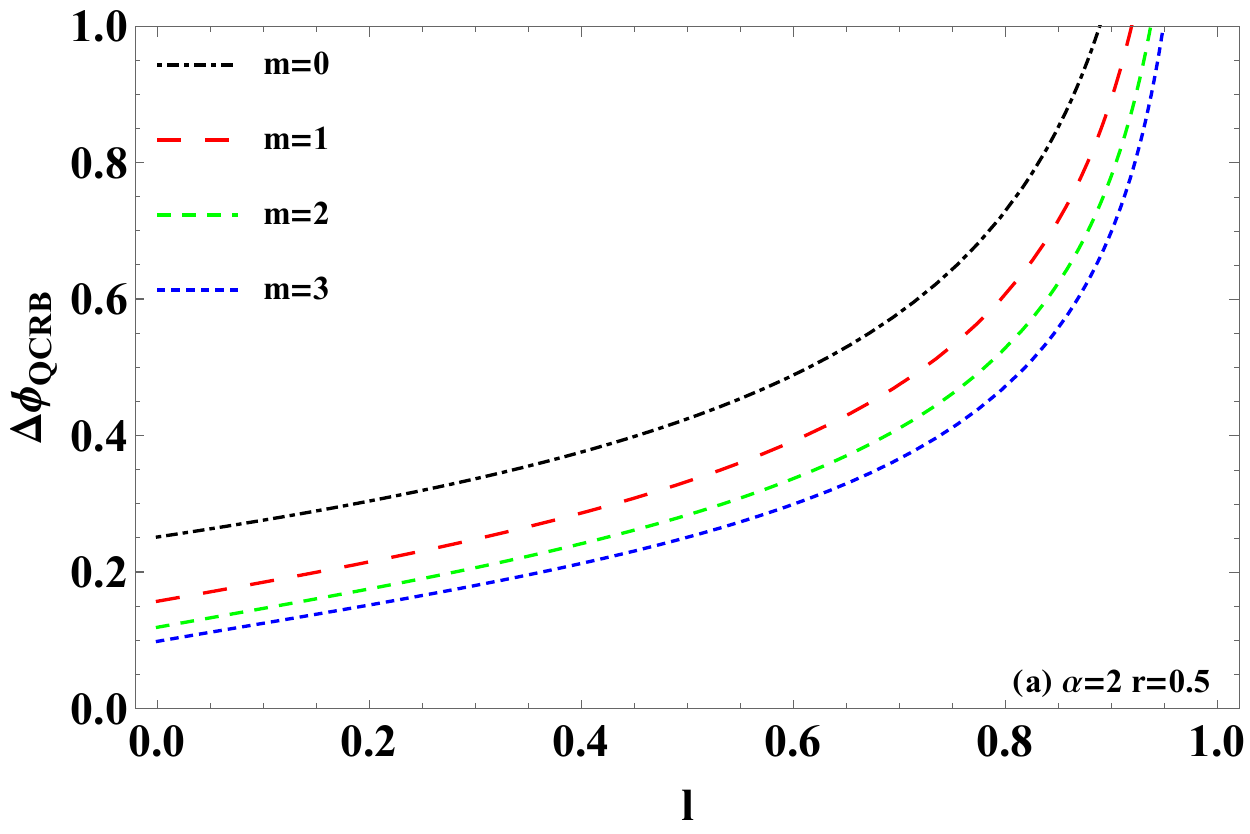}\\
\includegraphics[width=0.83\textwidth]{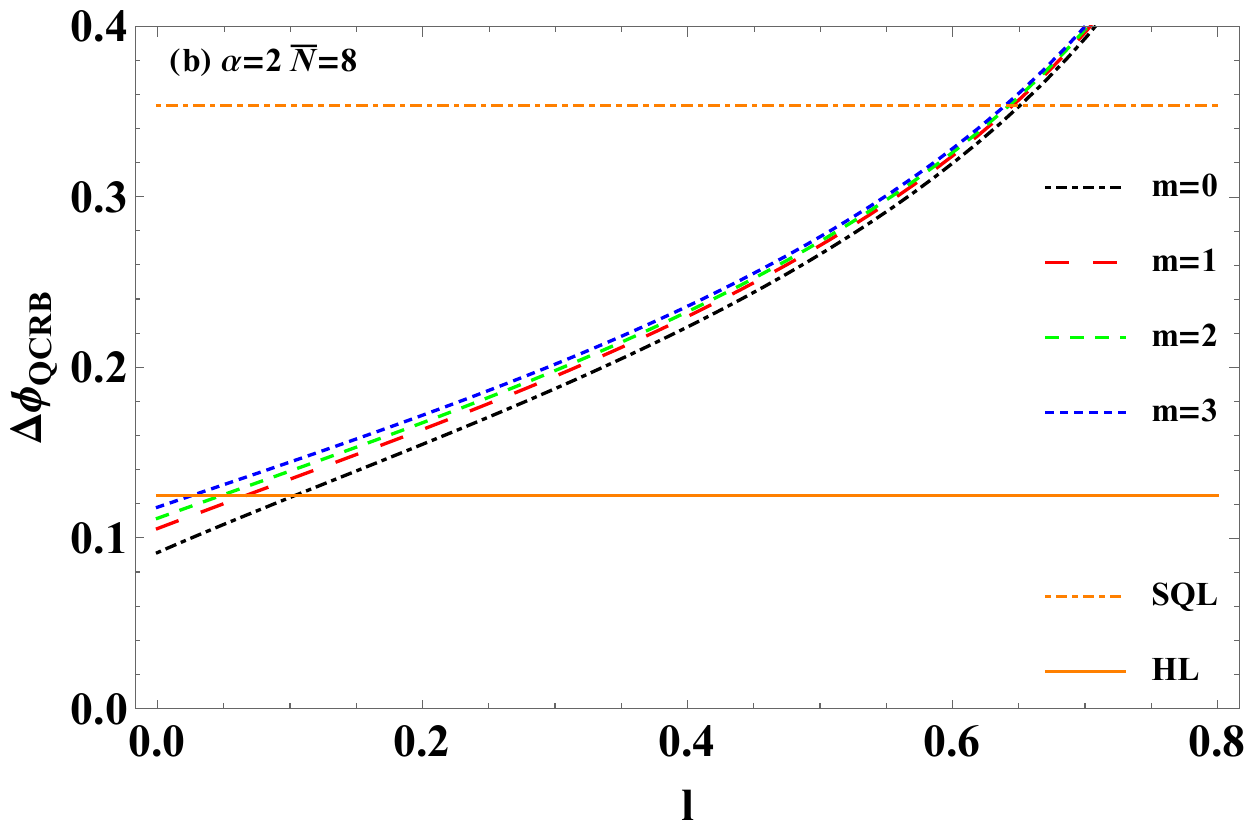}
\end{minipage}}
\caption{For the addition photon number $m=1,2,3$, and the coherent
amplitude $\protect \alpha =2$, the QCRB $\Delta \protect \phi _{QCRB}$ as a
function of loss rate $l$, (a) for given the squeezing parameter $r=0.5$,
and (b) for given the total average photon number $\bar{N}=8$, and compared
with the SQL and the HL.}
\end{figure}

In order to investigate the effect of photon loss on the QCRB $\Delta \phi
_{QCRB}$, we plot the variation of $\Delta \phi _{QCRB}$ with the loss rate $%
l$ in Fig. 8. It can be found that $\Delta \phi _{QCRB}$ deteriorates
gradually as $l$ increases. As can be seen in Fig. 8(a), $\Delta \phi
_{QCRB} $ can still be effectively improved by increasing $m$ in the case of
photon loss. This indicates that the robustness of the system to photon loss
is significantly enhanced by using photon addition operations. In Fig. 8(b),
it can be observed that $\Delta \phi _{QCRB}$ can effectively surpass the
SQL and even exceed the HL within a certain range of $l$, despite photon
loss. Additionally, the $\Delta \phi _{QCRB}$ is essentially the same for
different $m$ since $\bar{N}$ corresponding to the input resource energy is
fixed.
\begin{figure}[tbh]
\label{Fig9} \centering%
\subfigure{
\begin{minipage}[b]{0.5\textwidth}
\includegraphics[width=0.83\textwidth]{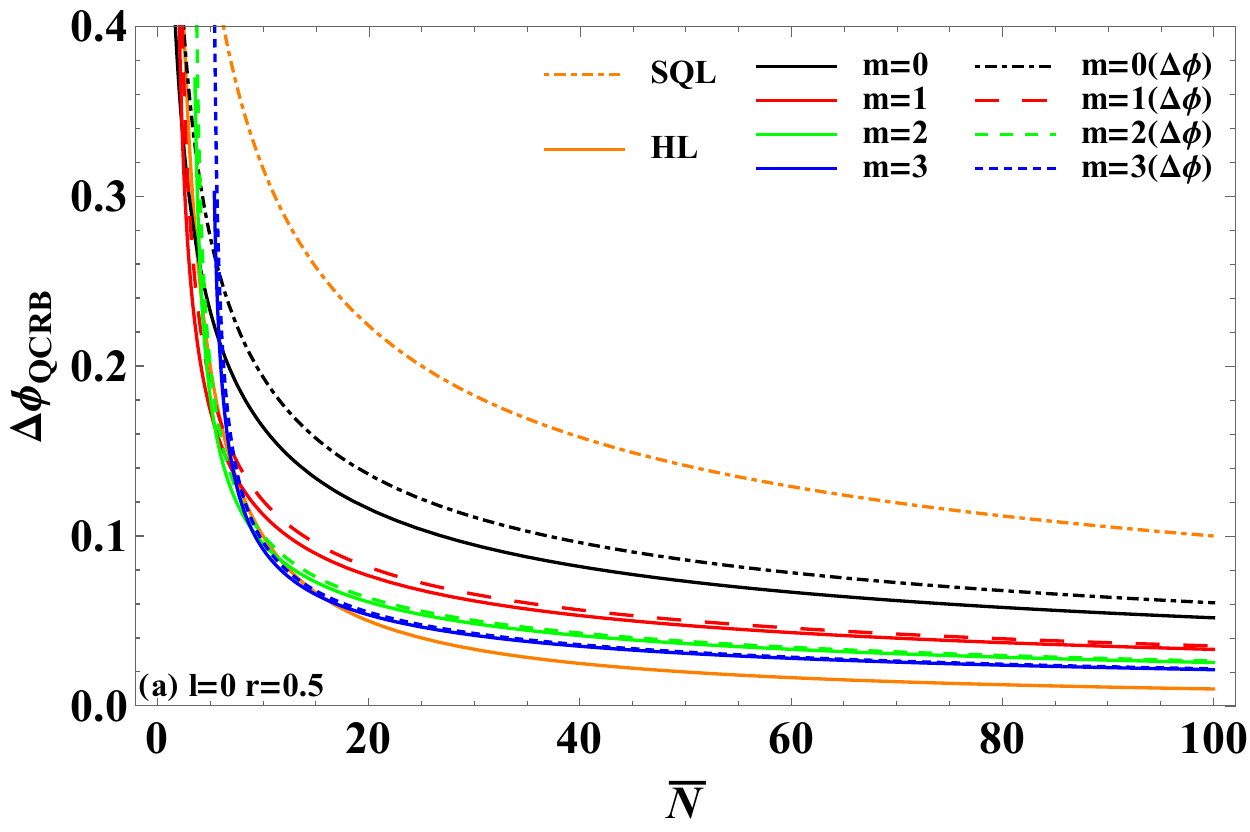}\\
\includegraphics[width=0.83\textwidth]{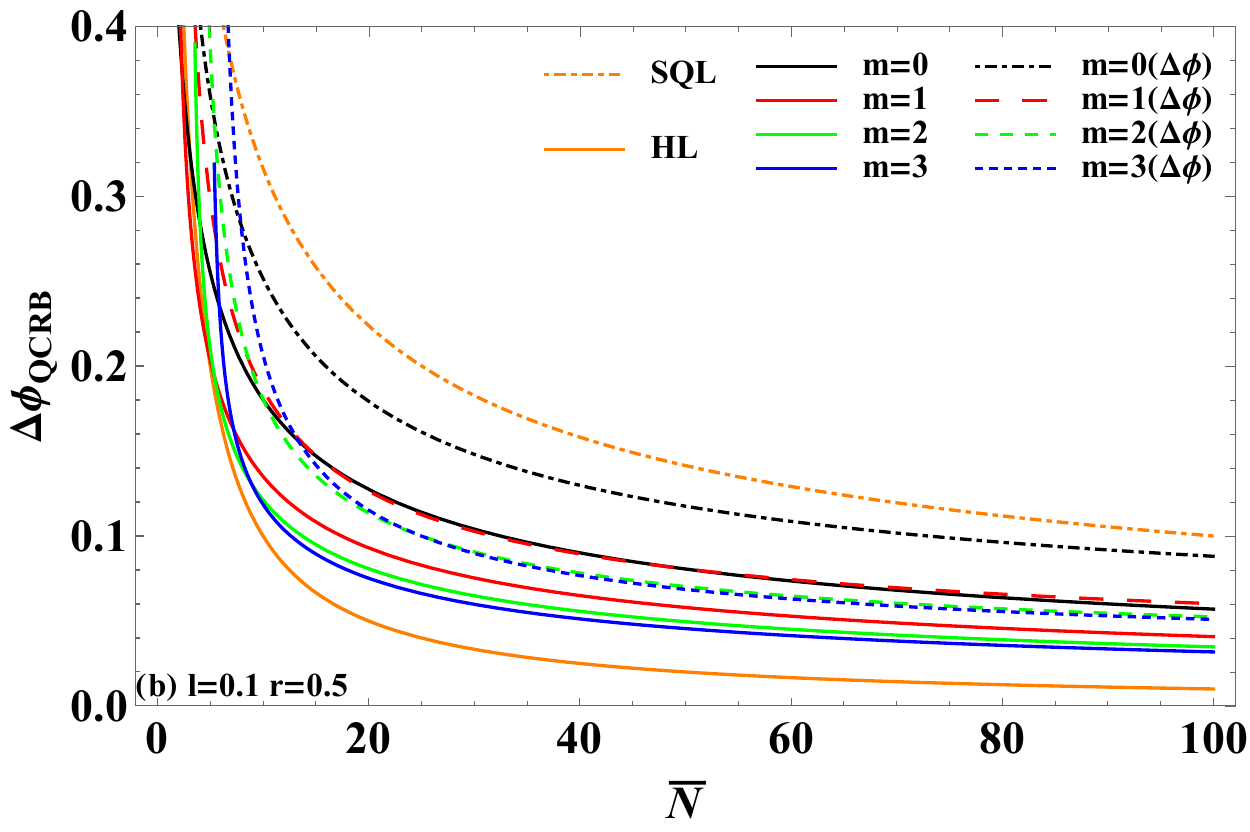}
\end{minipage}}
\caption{The QCRB $\Delta \protect \phi _{QCRB}$\ as a function of the total
average photon number $\overline{N}$ for the addition photon number $%
m=0,1,2,3$, and fixed the squeezing parameter $r=0.5$, (a) in the ideal case
of $l=0$, and (b) in the photon loss case of $l=0.1$. The SQL, the HL and
the variation of phase sensitivity $\Delta \protect \phi $ with respect to $%
\overline{N}$ (for $l=0$ and $l=0.1$, fixed the phase shift $\protect \phi %
=10^{-4}$ and optimization $\protect \phi $, respectively) are also plotted
for comparison.}
\end{figure}

As shown in Fig. 9, the variation of the QCRB $\Delta \phi _{QCRB}$ with
respect to the total average photon number $\bar{N}$ is demonstrated and
compared to the phase sensitivity $\Delta \phi $ as well as to the SQL and
the HL. It can be clearly seen that for both the ideal ($l=0$) and photon
loss ($l=0.1$) cases, $\Delta \phi _{QCRB}$ and $\Delta \phi $ can be
improved with the increase of $\bar{N}$ and $m$ when the fixed squeezing
parameter $r=0.5$. As illustrated in Fig. 9(a), under ideal condition, $%
\Delta \phi _{QCRB}$ and $\Delta \phi $ can evidently surpass the SQL and
even approach the HL. Additionally, as $m$ increases, $\Delta \phi $
progressively converges towards $\Delta \phi _{QCRB}$. Thus, in an ideal
scenario, optimal measurement for CS mixed PASVS input MZI scheme can be
achieved by employing parity detection. Fig. 9(b) illustrates that despite
photon loss, $\Delta \phi _{QCRB}$ and $\Delta \phi $ can still effectively
break through the SQL, and can be significantly improved by increasing $\bar{%
N}$ and utilizing photon addition operations.

\section{The QFI improved by Kerr nonlinear phase shifter}

Theoretically, the Kerr nonlinear transformation can substantially enhance
measurement accuracy compared to the phase encoding process using linear
transformation. Thus, this section further investigates the QFI in the case
of replacing the linear phase shifter $U\left( \phi \right) $ in the model
of Fig. 6 with a Kerr nonlinear phase shifter $U_{K}\left( \phi \right) $.
Building upon prior research methodologies for the QFI with linear phase
shift under both ideal and photon loss conditions, we further investigate
the QFI in the Kerr nonlinear phase shift case. The Kerr nonlinear phase
shifter operator is defined as $U_{K}\left( \phi \right) =\exp \left[ i\phi
\left( b^{\dagger }b\right) ^{2}\right] $, so according to Eq. (\ref{12})
and substituting $\left \vert \psi _{\phi }\right \rangle =U_{K}\left( \phi
\right) B_{1}\left \vert \psi \right \rangle _{in}$, the QFI for the Kerr
nonlinear case under the ideal condition can be derived as%
\begin{equation}
F_{Q}=4\left \langle \Delta ^{2}n_{b}^{2}\right \rangle =4\left[ \left
\langle n_{b}^{4}\right \rangle -\left \langle n_{b}^{2}\right \rangle ^{2}%
\right] .  \label{20}
\end{equation}%
Substituting $w=2$ and $w=4$ into Eq. (B3) to obtain $\left \langle
n_{b}^{2}\right \rangle $ and $\left \langle n_{b}^{4}\right \rangle $, and
further substituting them into Eq. (\ref{20}), the QFI for the ideal case of
Kerr nonlinear phase shift can be obtained.

Next, we investigate the QFI of photon loss for the Kerr nonlinear case. The
general form of the Kraus operator, incorporating the Kerr nonlinear phase
shift, is defined as follows \cite{46}:

\begin{equation}
\hat{\Pi}_{j}\left( \phi \right) =\sqrt{\frac{l^{j}}{j!}}e^{i\phi \left[
\left( b^{\dagger }b\right) ^{2}-2\mu _{1}b^{\dagger }bj-\mu _{2}j^{2}\right]
}\left( 1-l\right) ^{\frac{b^{\dagger }b}{2}}b^{j},  \label{21}
\end{equation}%
where $l$ is the loss rate, and the parameters $\mu _{1}=\mu _{2}=0$ or $-1$
correspond to photon loss occurring before or after the Kerr nonlinear phase
shifter. Referring to Eq. (\ref{15}), we derive (see Appendix C for
derivation):%
\begin{eqnarray}
F_{Q} &\leqslant &C_{Q}\left[ \left \vert \Psi _{S}\right \rangle ,\hat{\Pi}%
_{j}\left( \phi \right) \right]  \notag \\
&=&4\left[ K_{1}^{2}\left \langle \Delta ^{2}n_{b}^{2}\right \rangle
-K_{2}\left \langle n_{b}^{3}\right \rangle +K_{3}\left \langle
n_{b}^{2}\right \rangle \right.  \notag \\
&&\left. -K_{4}\left \langle n_{b}\right \rangle -K_{5}\left \langle
n_{b}^{2}\right \rangle \left \langle n_{b}\right \rangle -K_{6}\left
\langle n_{b}\right \rangle ^{2}\right] ,  \label{22}
\end{eqnarray}%
where $K_{i}$ ($i=1,2,3,4,5,6$) are detailed in Appendix C. To determine the
QFI $F_{Q}$ under photon loss for the Kerr nonlinear case, substitute the
optimal values of $\mu _{1}$ and $\mu _{2}$ ($\mu _{1opt}$ and $\mu _{2opt}$%
) into $C_{Q}$ to find $C_{Q\min }$. The specific expressions for $\mu
_{1opt}$ and $\mu _{2opt}$ are provided in Appendix C.
\begin{figure}[tbh]
\label{Fig10} \centering%
\subfigure{
\begin{minipage}[b]{0.5\textwidth}
\includegraphics[width=0.83\textwidth]{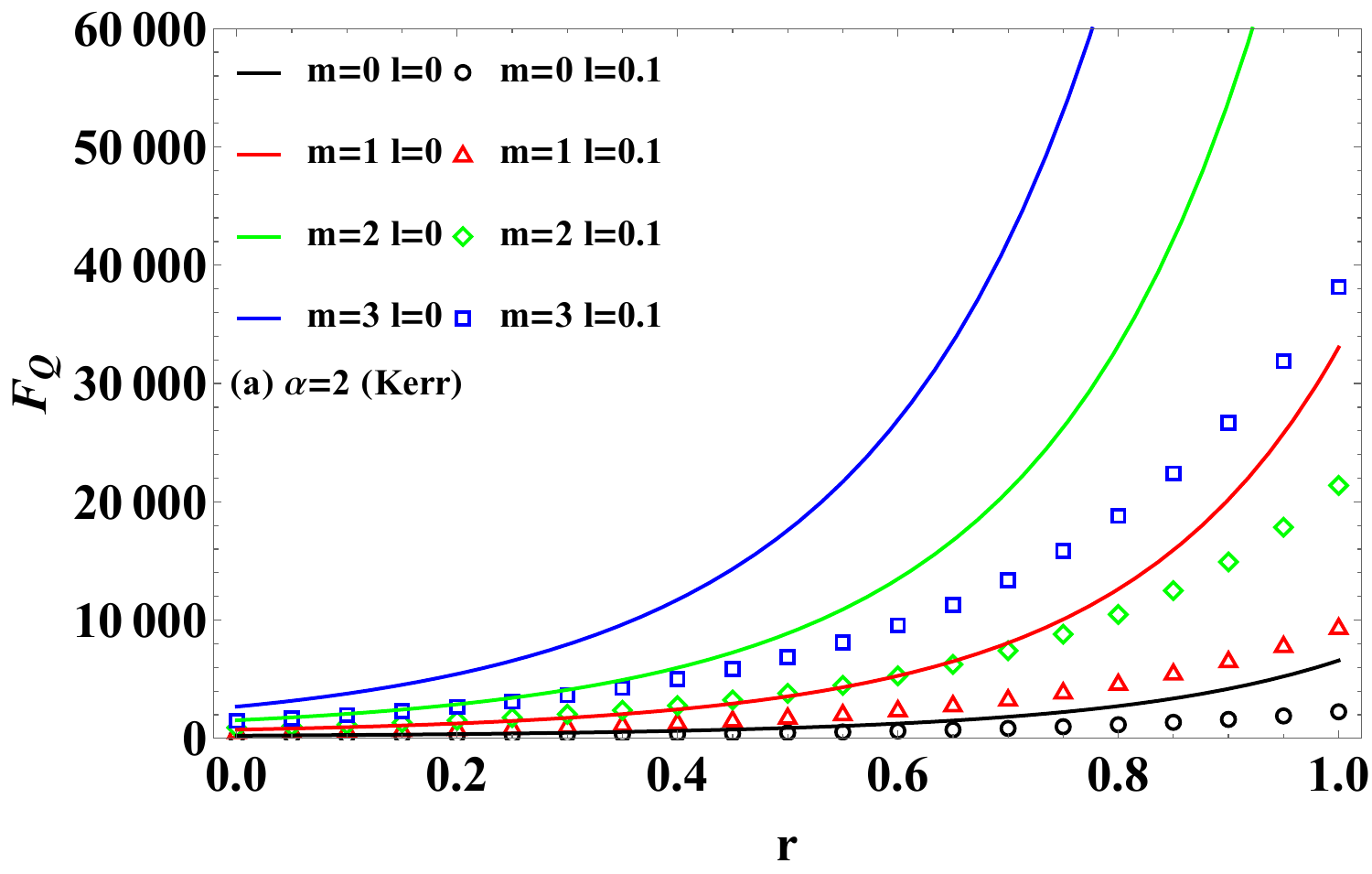}\\
\includegraphics[width=0.83\textwidth]{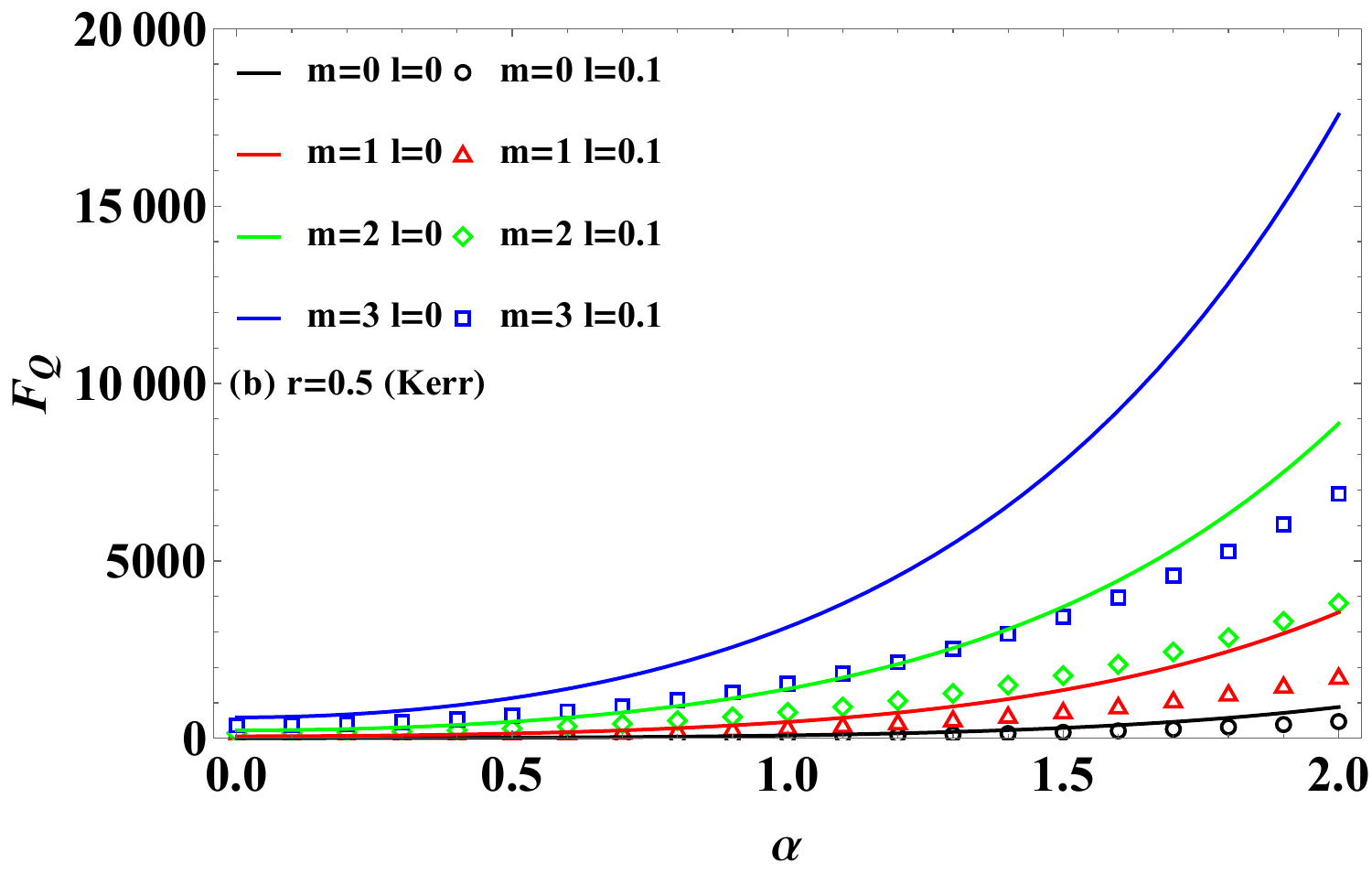}
\end{minipage}}
\caption{For the Kerr nonlinear phase shift case, and the addition photon
number $m=0,1,2,3$, in the ideal case of $l=0$ and in the photon loss case
of $l=0.1$, (a) the variation of QFI $F_{Q}$ with the squeezing parameter $r$
for given the coherent amplitude $\protect \alpha =2$, and (b) the variation
of $F_{Q}$ with $\protect \alpha $ for given $r=0.5$.}
\end{figure}

Fig. 10 illustrates the QFI $F_{Q}$ variation with the squeezing parameter $%
r $ and coherent amplitude $\alpha $ for the Kerr nonlinear case,
considering both ideal and photon loss scenarios. Clearly, $F_{Q}$ increases
with $r$, $\alpha $, and the addition photon number $m$. A comparison
between Fig. 7 and Fig. 10 indicates that the QFI exhibits a substantial
increase in the case of Kerr nonlinear phase shift relative to the linear
one. Hence, these results demonstrate that QFI and the phase measurement
accuracy of QCRB can be effectively increased by enhancing input resources
of $r$ and $\alpha $, utilizing photon addition operations, and the Kerr
nonlinear phase shifter.
\begin{figure}[tbh]
\label{Fig11} \centering%
\subfigure{
\begin{minipage}[b]{0.5\textwidth}
\includegraphics[width=0.83\textwidth]{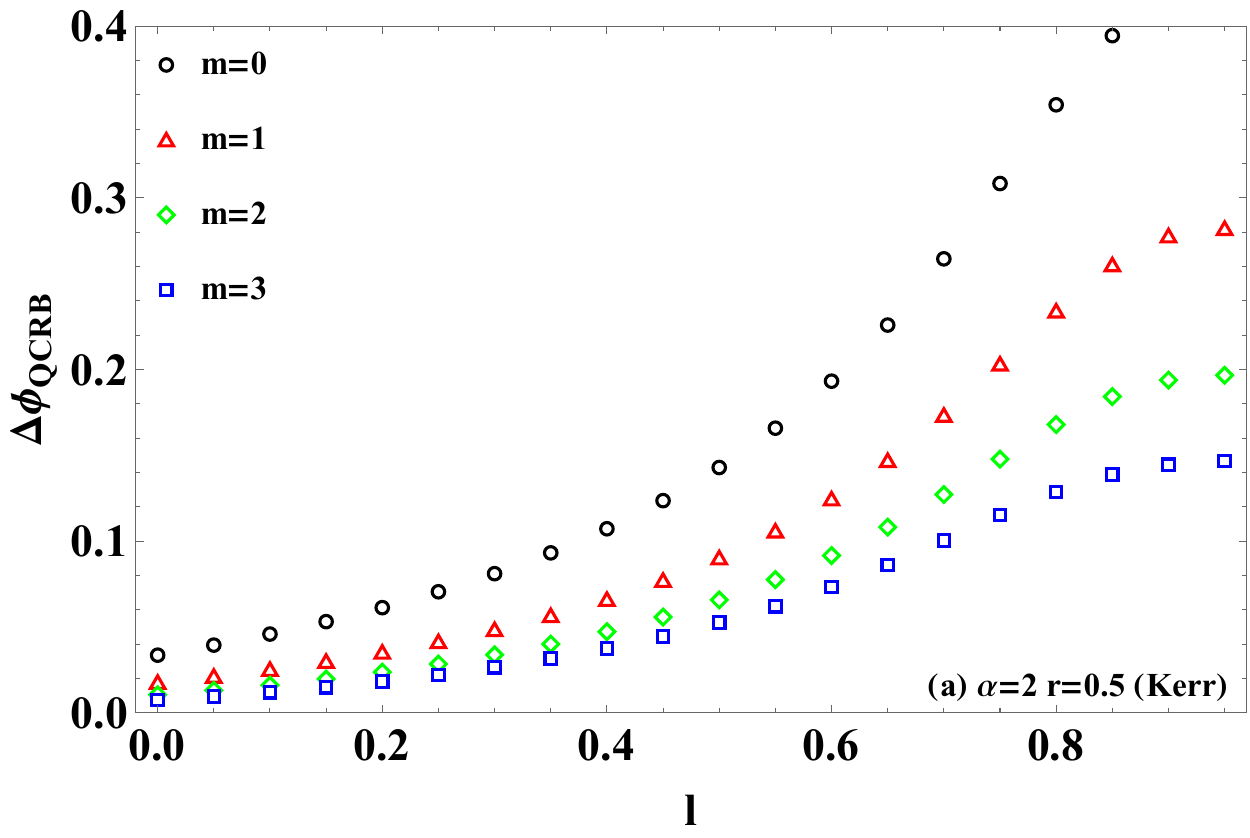}\\
\includegraphics[width=0.83\textwidth]{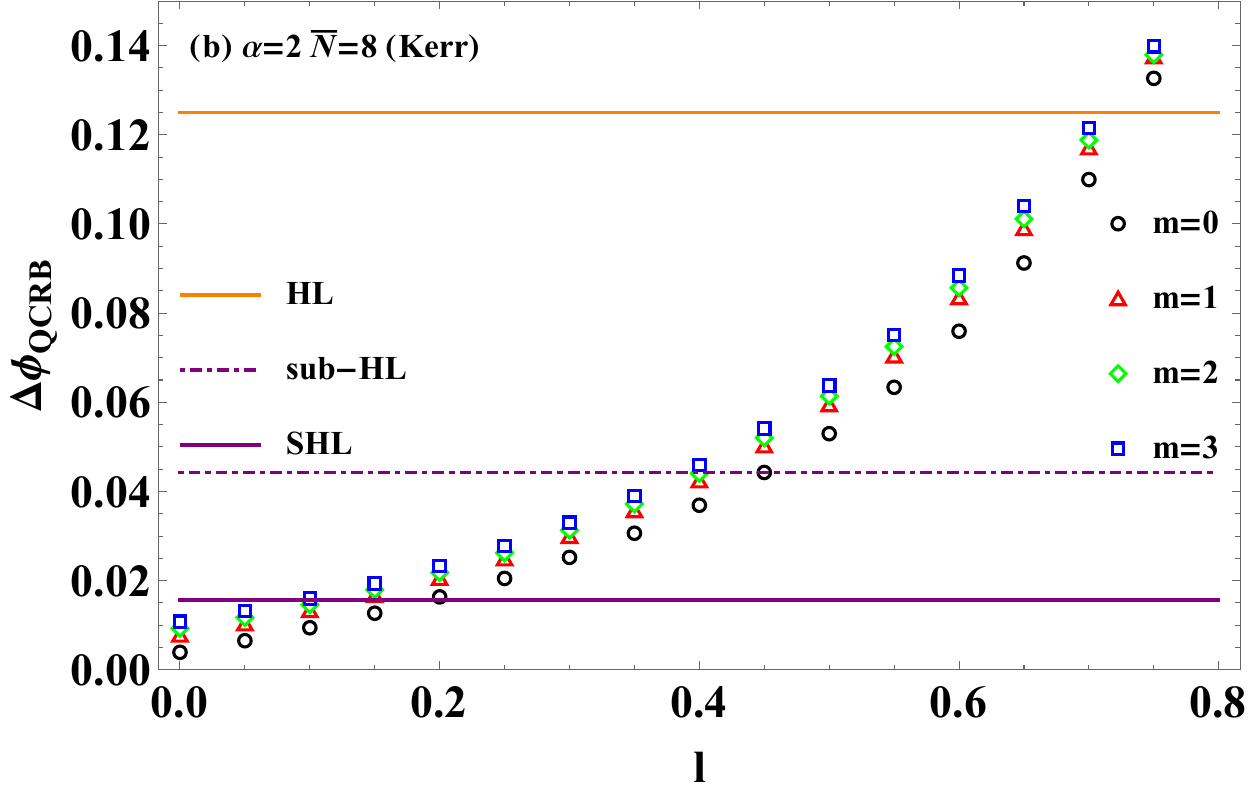}
\end{minipage}}
\caption{For the Kerr nonlinear phase shift case, the addition photon number
$m=1,2,3$, and the coherent amplitude $\protect \alpha =2$, the QCRB $\Delta
\protect \phi _{QCRB}$ as a function of loss rate $l$, (a) for given the
squeezing parameter $r=0.5$, and (b) for given the total average photon
number $\bar{N}=8$, and compared with the SQL and the HL.}
\end{figure}

For the Kerr nonlinear case, the variation of QCRB $\Delta \phi _{QCRB}$
with loss rate $l$ is shown in Fig. 11. By comparing Fig. 11(a) to Fig.
8(a), a similar conclusion can be obtained, i.e., although photon loss
worsens $\Delta \phi _{QCRB}$, the use of the photon addition operations
remains effective in improving $\Delta \phi _{QCRB}$, and $\Delta \phi
_{QCRB}$ improves as $m$ increases. $\Delta \phi _{QCRB}$ of Fig. 11 has a
significant improvement with respect to the linear phase shift case of Fig.
8. Moreover, it can be found from Fig. 11(b) that $\Delta \phi _{QCRB}$ can
break through the HL over a large range of $l$ and can break through the
sub-HL and even surpass the SHL over a certain range of $l$. This indicates
that the utilization of the Kerr nonlinear phase shifter can significantly
improve the QCRB and the robustness of the system to photon loss.
\begin{figure}[tbh]
\label{Fig12} \centering \includegraphics[width=0.83\columnwidth]{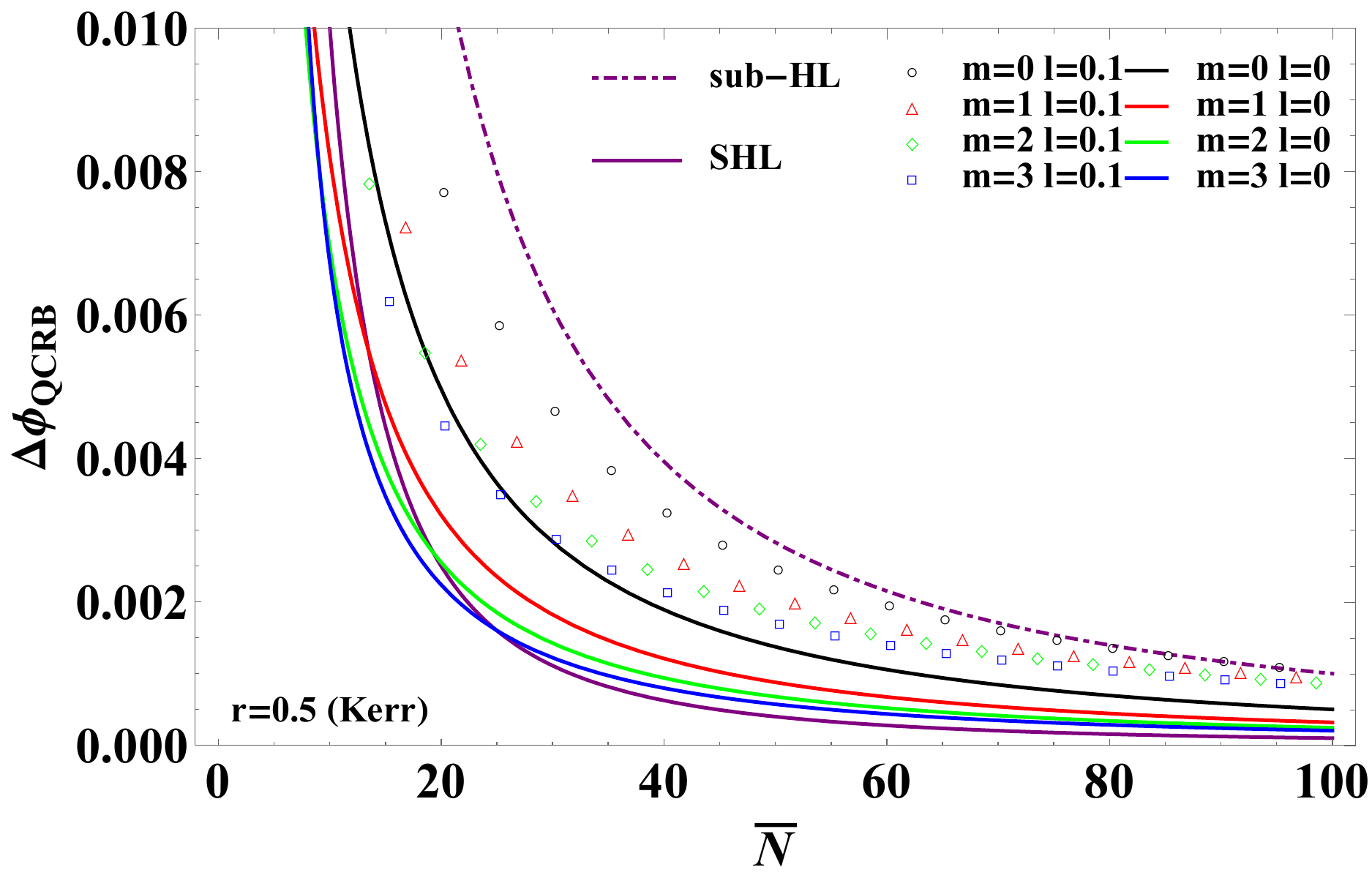}
\caption{For the Kerr nonlinear phase shift case, the addition photon number
$m=0,1,2,3$, and fixed the squeezing parameter $r=0.5$, in the ideal case of
$l=0$ and in the photon loss case of $l=0.1$, the variation of QCRB $\Delta
\protect \phi _{QCRB}$ with respect to the total average photon number $%
\overline{N}$. The HL, sub-HL, and SHL are also plotted for comparison.}
\end{figure}

To further reflect the improved effect of the Kerr nonlinear phase shift on
the accuracy, Fig. 12 depicts the variation of the QCRB $\Delta \phi _{QCRB}$
with respect to the total average photon number $\bar{N}$ and compares it
with the HL, sub-HL, and SHL. It is clear from the images that $\Delta \phi
_{QCRB}$ can be effectively improved for both the ideal ($l=0$) and photon
loss ($l=0.1$) cases by increasing $\bar{N}$ and $m$. Furthermore, in the
ideal case, $\Delta \phi _{QCRB}$ of the Kerr nonlinear phase shift can
break through the sub-HL to a large extent, and the photon addition
operations of $m=1,2,3$ can make $\Delta \phi _{QCRB}$ approach or even
exceed the SHL. In the case of $l=0.1$, $\Delta \phi _{QCRB}$ improved by
the photon addition operations can still effectively break through the
sub-HL. Thus, the utilization of both the photon addition operations and the
Kerr nonlinear phase shifter can significantly improve the measurement
accuracy.

\section{Conclusion}

We propose a scheme that utilizes CS and PASVS inputs in an MZI to improve
the precision of phase measurements. Our primary focus is on investigating
the phase sensitivity of parity detection and the QFI, both under ideal
conditions and in the presence of photon loss. The results indicate that
phase measurement accuracy is significantly improved by optimizing input
resources, i.e. increasing the squeezing parameter $r$, coherent amplitude $%
\alpha $, and total average photon number $\bar{N}$. Furthermore, the
comparison reveals that the non-Gaussian operations of photon addition
markedly enhance phase sensitivity and QFI. In the ideal case, the phase
sensitivity and QCRB improve with increasing the addition photon number $m$,
significantly surpassing the SQL and even reaching the HL. In addition, the
phase sensitivity can be better approximated to the QCRB by increasing $m$,
which indicates that the parity detection is an optimal measurement for the
phase estimation of the CS mixed PASVS inputs MZI. In the presence of photon
loss, phase sensitivity and QCRB can still be significantly enhanced by
increasing $m$, allowing the SQL to be surpassed across a relatively broad
range of loss rates $l$.

In order to achieve a more effective improvement in measurement accuracy
using the CS mixed PASVS input scheme, we further considered replacing the
linear phase shifter of MZI with a Kerr nonlinear phase shifter. Based on
the study of QFI for the Kerr nonlinear phase shift scenario, it is evident
that both QFI and QCRB can be substantially enhanced in both the ideal case
and the photon loss case by increasing the parameters $r$, $\alpha $, and $%
\bar{N}$, as well as optimizing the photon addition operation through an
increase in $m$. Comparing the linear phase shift case, it can be clearly
found that the Kerr nonlinear phase shift has a significant improvement on
the QFI and QCRB, and the QCRB at photon loss can break through the HL over
a wide range of the loss rate $l$, and can effectively exceed the sub-HL and
even surpass the SHL over a certain range. Our findings demonstrate that
employing photon addition operations and Kerr nonlinear phase shifter
effectively enhances phase measurement accuracy and robustness to photon
loss in practical applications. This research is pivotal for advancing
quantum precision measurement.

\begin{acknowledgments}
This work is supported by the National Natural Science Foundation of China
(Grants No. 11964013 and No. 12104195) and the Jiangxi Provincial Natural
Science Foundation (Grants No. 20242BAB26009 and 20232BAB211033), Jiangxi
Provincial Key Laboratory of Advanced Electronic Materials and Devices
(Grant No. 2024SSY03011), Jiangxi Civil-Military Integration Research
Institute (Grant No. 2024JXRH0Y07), as well as the Science and Technology
Project of Jiangxi Provincial Department of Science and Technology (Grant
No. GJJ2404102).
\end{acknowledgments}

\bigskip

\textbf{APPENDIX\ A : THE NORMAL ORDERING OF PARITY OPERATOR AND ITS AVERAGE
VALUE}

To facilitate the computation of the normal ordering form of the equivalent
operator for parity detection $\Pi _{b}^{loss}$ encompassing the entire
lossy MZI, we first present the Weyl ordering form representation of the
parity operator $\Pi _{b}$ in the ideal case \cite{60,61}, as follows

\begin{equation}
\Pi _{b}=\frac{\pi }{2}%
\begin{array}{c}
\colon \\
\colon%
\end{array}%
\delta \left( b\right) \delta \left( b^{\dag }\right)
\begin{array}{c}
\colon \\
\colon%
\end{array}%
,\   \tag{A1}
\end{equation}%
where$%
\begin{array}{c}
\colon \\
\colon%
\end{array}%
\bullet
\begin{array}{c}
\colon \\
\colon%
\end{array}%
$ is the Weyl ordering and $\delta \left( \cdot \right) $ denotes the delta
function. According to Eq. (\ref{8}), by employing the invariance of Weyl
ordering under similarity transformations \cite{62} and combining the linear
phase shift case with the operator transformation relation concerning the
lossy MZI, we obtain%
\begin{align}
\Pi _{b}^{loss}& =\left. _{f}\left \langle 0\right \vert U_{MZI}^{\dagger
}\Pi _{b}U_{MZI}\left \vert 0\right \rangle _{f}\right.  \notag \\
& =\frac{\pi }{2}\left. _{f}\left \langle 0\right \vert U_{MZI}^{\dagger }%
\begin{array}{c}
\colon \\
\colon%
\end{array}%
\delta \left( b\right) \delta \left( b^{\dag }\right)
\begin{array}{c}
\colon \\
\colon%
\end{array}%
U_{MZI}\left \vert 0\right \rangle _{f}\right.  \notag \\
& =\frac{\pi }{2}\left. _{f}\left \langle 0\right \vert
\begin{array}{c}
\colon \\
\colon%
\end{array}%
U_{MZI}^{\dagger }\delta \left( b\right) \delta \left( b^{\dag }\right)
U_{MZI}%
\begin{array}{c}
\colon \\
\colon%
\end{array}%
\left \vert 0\right \rangle _{f}\right. ,  \tag{A2}
\end{align}%
where $U_{MZI}=B_{2}U\left( \phi \right) B_{f}B_{1}$ represents the
equivalent operator of lossy MZI, considering photon loss in the $b$-mode.
By further utilizing the transformation relation of $U_{MZI}$ (substituting
the transformation relation of $U\left( \phi \right) $ as well as Eqs. (\ref%
{1})-(\ref{2}) and (\ref{7})), we can derive

\begin{align}
\Pi _{b}^{loss}& =\frac{\pi }{2}\left. _{f}\left \langle 0\right \vert
\begin{array}{c}
\colon \\
\colon%
\end{array}%
\delta \left \{ x_{1}a+x_{2}b+x_{3}b_{f}\right \} \right.  \notag \\
& \times \delta \left \{ x_{1}^{\ast }a^{\dagger }+x_{2}^{\ast }b^{\dagger
}+x_{3}^{\ast }b_{f}^{\dagger }\right \}
\begin{array}{c}
\colon \\
\colon%
\end{array}%
\left \vert 0\right \rangle _{f},  \tag{A3}
\end{align}%
where%
\begin{align}
x_{1}=& \frac{i}{2}\left( 1-e^{i\phi }\sqrt{1-l}\right) ,  \notag \\
x_{2}=& \frac{1}{2}\left( 1+e^{i\phi }\sqrt{1-l}\right) ,  \notag \\
x_{3}=& e^{i\phi }\sqrt{\frac{l}{2}}.  \tag{A4}
\end{align}%
By using the normal ordering form of the associated Wigner operator, i.e.,
\begin{align}
\Delta _{a}\left( \alpha \right) & =\frac{1}{\pi }\colon \exp \left[
-2\left( a-\alpha \right) \left( a^{\dagger }-\alpha ^{\ast }\right) \right]
\colon ,  \notag \\
\Delta _{b}\left( \beta \right) & =\frac{1}{\pi }\colon \exp \left[ -2\left(
b-\beta \right) \left( b^{\dagger }-\beta ^{\ast }\right) \right] \colon ,
\notag \\
\Delta _{b_{f}}\left( \gamma _{f}\right) & =\frac{1}{\pi }\colon \exp \left[
-2\left( b_{f}-\gamma _{f}\right) \left( b_{f}^{\dagger }-\gamma _{f}^{\ast
}\right) \right] \colon ,  \tag{A5}
\end{align}%
the classical correspondence of the Weyl ordering operator can be obtained
as follows \cite{63}

\begin{align}
&
\begin{array}{c}
\colon  \\
\colon
\end{array}%
f\left( a,a^{\dag },b,b^{\dag },b_{f},b_{f}^{\dag }\right)
\begin{array}{c}
\colon  \\
\colon
\end{array}
\notag \\
& =8\int d^{2}\alpha d^{2}\beta d^{2}\gamma _{f}f\left( \alpha ,\alpha
^{\ast },\beta ,\beta ^{\ast },\gamma _{f},\gamma _{f}^{\ast }\right)
\notag \\
& \times \Delta _{a}\left( \alpha \right) \Delta _{b}\left( \beta \right)
\Delta _{b_{f}}\left( \gamma _{f}\right) ,\   \tag{A6}
\end{align}%
and by combining the integration within an ordered product (IWOP) technique
with\bigskip \ the following integral formula

\begin{equation}
\int \frac{d^{2}z}{\pi }e^{\zeta \left \vert z\right \vert ^{2}+\xi z+\eta
z^{\ast }+fz^{2}+gz^{\ast 2}}=\frac{e^{\frac{-\zeta \xi \eta +\xi ^{2}g+\eta
^{2}f}{\zeta ^{2}-4fg}}}{\sqrt{\zeta ^{2}-4fg}},  \tag{A7}
\end{equation}%
one can further compute to obtain the normal ordering form of $\Pi
_{b}^{loss}$ as follows

\begin{equation}
\Pi _{b}^{loss}=\colon \exp \left[ X_{1}a^{\dagger }a+X_{2}b^{\dagger
}b+X_{3}a^{\dagger }b+X_{3}^{\ast }ab^{\dagger }\right] \colon ,  \tag{A8}
\end{equation}%
where%
\begin{align}
X_{1}& =\frac{l-2}{2}+\sqrt{1-l}\cos \phi ,  \notag \\
X_{2}& =\frac{l-2}{2}-\sqrt{1-l}\cos \phi ,  \notag \\
X_{3}& =\frac{il}{2}-\sqrt{1-l}\sin \phi .  \tag{A9}
\end{align}

To facilitate the use of the normal ordering of the parity operator $\Pi
_{b}^{loss}$ and compute its average value with respect to the input state,
we employ the coherent state representation. The input state can be
expressed as%
\begin{align}
\left \vert \psi \right \rangle _{in}& =\left \vert \alpha \right \rangle
_{a}\otimes \left \vert r,m\right \rangle _{b}  \notag \\
& =\frac{1}{\sqrt{P_{m}\cosh r}}\frac{\partial ^{m}}{\partial \tau ^{m}}\int
\frac{d^{2}\beta }{\pi }  \notag \\
& \times \left. \exp \left[ -\frac{\left \vert \beta \right \vert ^{2}}{2}%
+\tau \beta ^{\ast }-\frac{\tanh r}{2}\beta ^{\ast 2}\right] \right \vert
_{\tau =0}\left \vert \alpha \right \rangle _{a}\left \vert \beta \right
\rangle _{b},  \tag{A10}
\end{align}%
where $\left \vert \beta \right \rangle _{b}$ is a coherent state used to
represent the PASVS input on the $b$-mode. Using Eqs. (A8) and (A10) and
referring to Eq. (A7), we can calculate the average value of the parity
operator for under photon loss:%
\begin{align}
\left \langle \Pi _{b}\right \rangle & =\left. _{in}\left \langle \psi
\right \vert \Pi _{b}^{loss}\left \vert \psi \right \rangle _{in}\right.
\notag \\
& =\frac{\exp \left[ A_{1}\alpha ^{2}\right] }{P_{m}\sqrt{\bar{A}}\cosh r}
\notag \\
& \times \left. \frac{\partial ^{2m}}{\partial t^{m}\partial \tau ^{m}}\exp %
\left[ \frac{A_{2}+A_{3}\alpha }{\bar{A}}\right] \right \vert _{t=\tau =0},
\tag{A11}
\end{align}%
where%
\begin{align}
\bar{A}& =1-\tanh ^{2}r\left( 1+X_{2}\right) ^{2},  \notag \\
A_{1}& =X_{1}+\frac{\tanh ^{2}r\left( 1+X_{2}\right) \left \vert X_{3}\right
\vert ^{2}}{\bar{A}}-\frac{\tanh r}{2\bar{A}}\left( X_{3}^{2}+X_{3}^{\ast
2}\right) ,  \notag \\
A_{2}& =\left( 1+X_{2}\right) t\tau -\frac{\tanh r}{2}\left( 1+X_{2}\right)
^{2}\left( t^{2}+\tau ^{2}\right) ,  \notag \\
A_{3}& =X_{3}^{\ast }t+X_{3}\tau -\tanh r\left( 1+X_{2}\right) \left(
X_{3}t+X_{3}^{\ast }\tau \right) .  \tag{A12}
\end{align}%
In particular, for $l=0$, the result of Eq. (A11) represents the average
value of the parity operator in the ideal scenario.

\textbf{APPENDIX\ B : THE NORMAL ORDERING FOR OPERATOR IDENTITY AND ITS
AVERAGE VALUE}

This appendix derives the average of $n_{b}^{w}$ with respect to $%
\left
\vert \Psi \right \rangle _{S}$ in order to obtain an expression for
the simplicity of calculating the QFI. The normal ordering of $n_{b}^{w}$ is
obtained via the operator identity:

\begin{align}
n_{b}^{w}& =\frac{\partial ^{w}}{\partial x^{w}}\left. \exp \left[
xb^{\dagger }b\right] \right \vert _{x=0}  \notag \\
& =\frac{\partial ^{w}}{\partial x^{w}}\left. \colon \exp \left[ \left(
e^{x}-1\right) b^{\dagger }b\right] \colon \right \vert _{x=0}.  \tag{B1}
\end{align}%
Based on the coherent state representation of $\left \vert \psi
\right
\rangle _{in}$ (Eq. (A10)) and utilizing the transformation relation
for BS1 (Eq. (\ref{1})), the form of $\left \vert \Psi \right \rangle _{S}$
in the coherent state representation is given by%
\begin{align}
\left \vert \Psi \right \rangle _{S}& =\frac{1}{\sqrt{P_{m}\cosh r}}\frac{%
\partial ^{m}}{\partial \tau ^{m}}\int \frac{d^{2}\beta }{\pi }\exp \left[ -%
\frac{\left \vert \beta \right \vert ^{2}}{2}+\tau \beta ^{\ast }\right]
\notag \\
& \times \left. \exp \left[ -\frac{\tanh r}{2}\beta ^{\ast 2}\right] \right
\vert _{\tau =0}\left \vert \alpha _{1}\right \rangle _{a}\left \vert \beta
_{1}\right \rangle _{b},  \tag{B2}
\end{align}%
where $\alpha _{1}=\left( \alpha -i\beta \right) /\sqrt{2}$ and $\beta
_{1}=\left( \beta -i\alpha \right) /\sqrt{2}$. Using Eqs. (B1) and (B2)
along with the IWOP technique and the integral formula Eq. (A7), we obtain%
\begin{equation}
\left \langle n_{b}^{w}\right \rangle =D_{m,w}\left \{ E_{M}\right \} ,
\tag{B3}
\end{equation}%
where
\begin{align}
D_{m,w}\left \{ \cdot \right \} & =\frac{1}{P_{m}\cosh r}\frac{\partial
^{2m+w}}{\partial t^{m}\partial \tau ^{m}\partial x^{w}}\left. \left \{
\cdot \right \} \right \vert _{t=\tau =x=0},  \notag \\
E_{M}& =\frac{M_{1}}{\sqrt{M_{0}}}\exp \left[ \frac{2M_{2}t-M_{2}^{2}\tanh r%
}{2M_{0}}\right]  \notag \\
& \times \exp \left[ -\frac{\left( s+1\right) ^{2}t^{2}\tanh r}{2M_{0}}%
\right] ,  \tag{B4}
\end{align}%
and

\begin{align}
M_{0}& =1-\left( s+1\right) ^{2}\tanh ^{2}r,  \notag \\
M_{1}& =\exp \left[ s\alpha \left( \alpha +i\tau +\frac{s\alpha \tanh r}{2}%
\right) \right] ,  \notag \\
M_{2}& =-is\alpha +\left( e^{x}-s\right) \left( \tau -is\alpha \tanh
r\right) ,  \tag{B5}
\end{align}%
where $s=\frac{1}{2}\left( e^{x}-1\right) $.

\textbf{APPENDIX\ C : }$C_{Q}$\textbf{\ FOR THE KERR NONLINEAR PHASE SHIFT
CASE}

In this appendix, we derive $C_{Q}$ for the Kerr nonlinear phase shift and
its specific expression to obtain the QFI \thinspace $F_{Q}$ under photon
loss condition. To compute $C_{Q}$ using Eqs. (\ref{15})-(\ref{17}) and (\ref%
{21}) according to the method in Refs. \cite{46} and \cite{59}, we first
obtain the normal ordering form of $\left( 1-l\right) ^{n_{b}}n_{b}^{q}$ by
utilizing the operator identity from Eq. (B1) as follows%
\begin{align}
\left( 1-l\right) ^{n_{b}}n_{b}^{q}& =\eta ^{n_{b}}n_{b}^{q}  \notag \\
& =\frac{\partial ^{q}}{\partial x^{q}}\left. \exp \left[ n_{b}\ln \eta %
\right] \exp \left[ n_{b}x\right] \right \vert _{x=0}  \notag \\
& =\colon \left. \frac{\partial ^{q}}{\partial x^{q}}e^{\left( \eta
e^{x}-1\right) b^{\dagger }b}\right \vert _{x=0}\colon ,  \tag{C1}
\end{align}%
where for simplicity, we set $\eta =1-l$. Based on this equation and further
utilizing the IWOP technique, the following summation can be computed for
the operators $S_{q,p}$ associated with the Hermitian operators $H_{1,2}$
from Eqs. (\ref{16}) and (\ref{17})\ to obtain the generalized equation
about $n_{b}$ expressed in terms of the partial differential operator $%
D_{q,p}=\frac{\partial ^{q+p}}{\partial x^{q}\partial y^{p}}\left. \left[
\cdot \right] \right \vert _{x=y=0}$ as%
\begin{align}
S_{q,p}& =\sum_{j=0}^{\infty }\frac{\left( 1-\eta \right) ^{j}}{j!}%
j^{p}b^{\dagger j}\eta ^{n_{b}}n_{b}^{q}b^{j}  \notag \\
& =\sum_{j=0}^{\infty }\frac{\left( 1-\eta \right) ^{j}}{j!}j^{p}\colon
\left. \left( b^{\dagger }b\right) ^{j}\frac{\partial ^{q}}{\partial x^{q}}%
e^{\left( \eta e^{x}-1\right) b^{\dagger }b}\right \vert _{x=0}\colon  \notag
\\
& =\colon \sum_{j=0}^{\infty }\frac{\left[ \left( 1-\eta \right) b^{\dagger
}b\right] ^{j}}{j!}\frac{\partial ^{q+p}}{\partial x^{q}\partial y^{p}}%
\left. e^{\left( \eta e^{x}-1\right) b^{\dagger }b+yj}\right \vert
_{x=y=0}\colon  \notag \\
& =\colon \frac{\partial ^{q+p}}{\partial x^{q}\partial y^{p}}\left. e^{%
\left[ \eta e^{x}+\left( 1-\eta \right) e^{y}-1\right] b^{\dagger }b}\right
\vert _{x=y=0}\colon  \notag \\
& =\frac{\partial ^{q+p}}{\partial x^{q}\partial y^{p}}\left. \left[ \eta
e^{x}+\left( 1-\eta \right) e^{y}\right] ^{n_{b}}\right \vert _{x=y=0}.
\tag{C2}
\end{align}%
The final step in the above equation employs the operator identity from Eq.
(B1) for conversion.

According to Eq. (\ref{15}), by substituting the Kraus operator for the Kerr
nonlinear phase shift case (Eq. (\ref{21})) into Eqs. (\ref{16}) and (\ref%
{17}),\ and using Eq. (C2) one can further compute that%
\begin{align}
C_{Q}& =4[K_{1}^{2}\left \langle \Delta ^{2}n_{b}^{2}\right \rangle
-K_{2}\left \langle n_{b}^{3}\right \rangle +K_{3}\left \langle
n_{b}^{2}\right \rangle  \notag \\
& -K_{4}\left \langle n_{b}\right \rangle -K_{5}\left \langle
n_{b}^{2}\right \rangle \left \langle n_{b}\right \rangle -K_{6}\left
\langle n_{b}\right \rangle ^{2}],  \tag{C3}
\end{align}%
where%
\begin{align}
K_{1}& =\omega _{1}\eta ^{2}-2\omega _{2}\eta -\mu _{2},  \notag \\
K_{2}& =2\eta \left[ 3\omega _{1}^{2}\eta ^{3}-3\omega _{3}^{2}\eta
^{2}-\omega _{4}\eta +\omega _{5}\right] ,  \notag \\
K_{3}& =\eta \left[ 11\omega _{1}^{2}\eta ^{3}-2\omega _{6}^{2}\eta
^{2}+\omega _{7}\eta -4\omega _{1}\omega _{2}\right] ,  \notag \\
K_{4}& =\eta \omega _{1}^{2}\left( 6\eta ^{3}-12\eta ^{2}+7\eta -1\right) ,
\notag \\
K_{5}& =2\left( 1-\eta \right) \eta \omega _{1}K_{1},  \notag \\
K_{6}& =\left( 1-\eta \right) ^{2}\eta ^{2}\omega _{1}^{2},  \tag{C4}
\end{align}%
and%
\begin{align}
\omega _{1}& =1+2\mu _{1}-\mu _{2},  \notag \\
\omega _{2}& =\mu _{1}-\mu _{2},  \notag \\
\omega _{3}& =1+2\left( 3\mu _{1}-2\mu _{2}\right) +\left( 2\mu _{1}-\mu
_{2}\right) \left( 4\mu _{1}-3\mu _{2}\right) ,  \notag \\
\omega _{4}& =7\mu _{2}-6\mu _{1}+24\mu _{1}\mu _{2}-14\mu _{1}^{2}-9\mu
_{2}^{2},  \notag \\
\omega _{5}& =\mu _{2}\omega _{1}-2\omega _{2}^{2},  \notag \\
\omega _{6}& =9+40\mu _{1}-22\mu _{2}+44\mu _{1}^{2}-48\mu _{1}\mu
_{2}+13\mu _{2}^{2},  \notag \\
\omega _{7}& =7+40\mu _{1}-26\mu _{2}+52\mu _{1}^{2}-64\mu _{1}\mu
_{2}+19\mu _{2}^{2},  \tag{C5}
\end{align}%
where parameters $\mu _{1}$ and $\mu _{2}$ are optimizable to describe the
photon losses occurring before and after the phase shifter. In particular, $%
\mu _{1}=\mu _{2}=0$ or $-1$ represents photon losses occurring before or
after the phase shifter, respectively. Using Eqs. (C3)-(C5), optimizing
through $\partial C_{Q}/\partial \mu _{1}=\partial C_{Q}/\partial \mu _{2}=0$
to find the minimum value of $C_{Q}$ gives%
\begin{align}
\mu _{1opt}& =\frac{G_{2}G_{5}-G_{3}G_{4}}{G_{1}G_{4}-2\eta G_{2}^{2}},
\tag{C6} \\
\mu _{2opt}& =\frac{G_{1}G_{5}-2\eta G_{2}G_{3}}{G_{1}G_{4}-2\eta G_{2}^{2}},
\tag{C7}
\end{align}%
where%
\begin{align}
G_{1}& =2\left[ -\left( 1-\eta \right) \eta \left( \left \langle \Delta
^{2}n_{b}^{2}\right \rangle +2\left \langle n_{b}^{2}\right \rangle \left
\langle n_{b}\right \rangle -\left \langle n_{b}\right \rangle ^{2}\right)
\right.  \notag \\
& -\left( 6\eta ^{2}-6\eta +1\right) \left( \left \langle n_{b}^{3}\right
\rangle +\left \langle n_{b}\right \rangle \right)  \notag \\
& \left. +\left( 11\eta ^{2}-11\eta +2\right) \left \langle n_{b}^{2}\right
\rangle \right] ,  \tag{C8}
\end{align}%
\begin{align}
G_{2}& =\left( 1-\eta \right) ^{2}\left \langle \Delta ^{2}n_{b}^{2}\right
\rangle +3\left( 1-\eta \right) \left( 2\eta -1\right) \left \langle
n_{b}^{3}\right \rangle  \notag \\
& +\left( 11\eta ^{2}-13\eta +3\right) \left \langle n_{b}^{2}\right \rangle
-\left( 6\eta ^{2}-6\eta +1\right) \left \langle n_{b}\right \rangle  \notag
\\
& -\left( 1-\eta \right) \left( 2\eta -1\right) \left \langle
n_{b}^{2}\right \rangle \left \langle n_{b}\right \rangle +\eta \left(
1-\eta \right) \left \langle n_{b}\right \rangle ^{2},  \tag{C9}
\end{align}%
\begin{align}
G_{3}& =\eta ^{2}\left \langle \Delta ^{2}n_{b}^{2}\right \rangle -3\eta
\left( 2\eta -1\right) \left \langle n_{b}^{3}\right \rangle  \notag \\
& +\left( 11\eta ^{2}-9\eta +1\right) \left \langle n_{b}^{2}\right \rangle
-\left( 6\eta ^{2}-6\eta +1\right) \left \langle n_{b}\right \rangle  \notag
\\
& +\eta \left( 2\eta -1\right) \left \langle n_{b}^{2}\right \rangle \left
\langle n_{b}\right \rangle +\eta \left( 1-\eta \right) \left \langle
n_{b}\right \rangle ^{2},  \tag{C10}
\end{align}%
\begin{align}
G_{4}& =-\left( 1-\eta \right) ^{3}\left \langle \Delta ^{2}n_{b}^{2}\right
\rangle -6\eta \left( 1-\eta \right) ^{2}\left \langle n_{b}^{3}\right
\rangle  \notag \\
& -\eta \left( 1-\eta \right) \left( 11\eta -4\right) \left \langle
n_{b}^{2}\right \rangle -\eta \left( 6\eta ^{2}-6\eta +1\right) \left
\langle n_{b}\right \rangle  \notag \\
& +2\eta \left( 1-\eta \right) ^{2}\left \langle n_{b}^{2}\right \rangle
\left \langle n_{b}\right \rangle +\eta ^{2}\left( 1-\eta \right) \left
\langle n_{b}\right \rangle ^{2},  \tag{C11}
\end{align}%
\begin{align}
G_{5}& =\eta \left[ -\eta \left( 1-\eta \right) \left( \left \langle \Delta
^{2}n_{b}^{2}\right \rangle -\left \langle n_{b}\right \rangle ^{2}\right)
\right.  \notag \\
& -\left( 6\eta ^{2}-6\eta +1\right) \left( \left \langle n_{b}^{3}\right
\rangle +\left \langle n_{b}\right \rangle \right)  \notag \\
& +\left( 11\eta ^{2}-11\eta +2\right) \left \langle n_{b}^{2}\right \rangle
\notag \\
& \left. +\left( 2\eta ^{2}-2\eta +1\right) \left \langle n_{b}^{2}\right
\rangle \left \langle n_{b}\right \rangle \right] ,  \tag{C12}
\end{align}%
and substituting these into Eq. (C3), along with using Eq. (B3) for $%
w=1,2,3,4$, results in $F_{Q}=C_{Q\min }$.

\end{document}